\documentclass[twocolumn]{emulateapj}
\usepackage[bookmarks,bookmarksopen,colorlinks,linkcolor={blue},citecolor={blue},
  urlcolor={red}]{hyperref}
\usepackage{amsmath}
\usepackage{natbib}
\usepackage{color}
\usepackage{graphics}
\usepackage{enumerate}
\usepackage{setspace}
\def\ryan[#1]{\noindent{\textbf{\textcolor{green}{#1}}}}
\def\kristen[#1]{\noindent{\textbf{\textcolor{red}{#1}}}}
\def\rene[#1]{\noindent{\textbf{\textcolor{blue}{#1}}}}
\newcommand{\spirou}[1]{\emph{SPIRou}#1}
\newcommand{\prob}{\mathrm{P}}
\newcommand{\nobs}[1]{N$_{\mathrm{obs}}$#1}
\newcommand{\mps}[1]{m s$^{-1}$#1}
\defcitealias{berta15}{B15}
\defcitealias{weiss14}{WM14}

\shortauthors{Cloutier et al.}
\shorttitle{RV search for multiple planets in transiting M-dwarf systems}

\begin{document}
\title{On the radial velocity detection of additional planets in transiting,
slowly rotating M-dwarf systems: the case of GJ 1132}
\author{Ryan Cloutier\altaffilmark{1,2,3}}
\author{Ren\'{e} Doyon\altaffilmark{3}}
\author{Kristen Menou\altaffilmark{2,1}}
\author{Xavier Delfosse\altaffilmark{4,5}}
\author{Xavier Dumusque\altaffilmark{6}}
\author{\'{E}tienne Artigau\altaffilmark{3}}
\altaffiltext{1}{Dept. of Astronomy \& Astrophysics, University
of Toronto. 50 St. George Street, Toronto, Ontario, Canada,
M5S 3H4}
\altaffiltext{2}{Centre for Planetary Sciences,
Dept. of Physical \& Environmental Sciences, University of
Toronto Scarborough. 1265 Military Trail, Toronto, Ontario, Canada,
M1C 1A4}
\altaffiltext{3}{Institut de recherche sur les exoplan\`{e}tes,
D\'{e}partement de physique, Universit\'{e} de Montr\'{e}al.
2900 boul. Édouard-Montpetit, Montr\'{e}al Quebec, Canada,
H3T 1J4}
\altaffiltext{4}{Universit\'{e} Grenoble Alpes, IPAG, F-38000 Grenoble, France}
\altaffiltext{5}{CNRS, IPAG, F-38000 Grenoble, France}
\altaffiltext{6}{Observatoire Astronomique de l'Universit\'{e} de Gen\`{e}ve, 51 
Chemin des Maillettes, 1290 Sauverny, Suisse}

\begin{abstract}
M-dwarfs are known to 
commonly host high-multiplicity planetary systems. Therefore M-dwarf planetary systems 
with a known transiting planet are expected to contain additional small planets 
($r_p \le 4$ R$_{\oplus}$, $m_p \lesssim 20$ M$_{\oplus}$) that are not seen in transit. 
In this study we investigate the effort required to detect such planets 
using precision velocimetry around 
the sizable subset of M-dwarfs which are slowly rotating ($P_{\mathrm{rot}} \gtrsim 40$ 
days) and hence more likely to be inactive. We focus on the test case of GJ 1132.
Specifically, we perform a suite of 
Monte-Carlo simulations of the star's radial velocity signal featuring astrophysical
contributions from stellar jitter due to rotationally modulated active regions and  
keplarian signals from the known transiting planet and hypothetical
additional planets not seen in transit. 
We then compute the detection completeness of non-transiting planets around GJ 1132
and consequently estimate the number of RV measurements required to detect those planets. 
We show that with 1 \mps{} precision per measurement, only $\sim 50$ measurements
are required to achieve a 50\% detection 
completeness to all non-transiting planets in the system and to planets
which are potentially habitable. 
Throughout we advocate the use of Gaussian process regression as an
effective tool for mitigating the effects of stellar 
jitter including stars with high activity.  
Given that GJ 1132 is representative of a large population of slowly rotating 
M-dwarfs, we conclude with a discussion of how our results may be extended to other
systems with known transiting planets such as those which will be discovered 
with \emph{TESS}.
\end{abstract}

\section{Introduction}
One major endeavour in current exoplanet research is the characterization of 
exoplanetary 
atmospheres through spectroscopic investigation. Indeed such efforts have already 
been performed from the ground as well as from space-based observatories including 
the \emph{Hubble Space Telescope} and \emph{Spitzer Space Telescope}, 
on primarily \emph{transiting} giant planets 
\citep{bean13, crouzet14, mccullough14} and even on planets down to $\sim 3$ Earth 
radii \citep{kreidberg14a}. High-dispersion spectroscopy may also be used to 
characterize
\emph{non-transiting} exoplanet atmospheres including measurements of orbital 
inclinations and planetary rotation velocities \citep{snellen13}. 
With upcoming technological advances on-board the 
\emph{James Webb Space Telescope}, researchers in the field are pushing the 
boundary towards the characterization of exoplanetary atmospheres around planets 
with bulk Earth-like compositions \citep{beichman14}. Such experiments require 
nearby planetary systems and favors hot planets around 
small host stars wherein the atmospheric transmission signal is maximized 
\citep{brown01}. Therefore the most favorable targets for characterizing potential
Earth-like exoplanetary atmospheres are around nearby M-dwarf stars.

Numerous current and upcoming transiting exoplanet missions (e.g. K2,
TESS, CHEOPS) will discover nearby M-dwarf transiting planetary
systems which are sufficiently bright and are therefore amenable to atmospheric
characterization. For example, the upcoming 
\emph{Transiting Exoplanet Survey Satellite}
mission \citep{ricker14} is expected to detect
$\sim 980$ transiting planets around M-dwarfs ($T_{\mathrm{eff}} \le 3800$ K)
including $\sim 50$ \emph{bright} M-dwarfs with $J < 9.5$ \citep{sullivan15}. 
Given the high frequency of Earth to mini-Neptune sized planets around early-mid 
M-dwarfs 
\citep[$\sim 0.36-2.5$ planets per M-dwarf;][]{bonfils13, dressing15a, gaidos16}, 
including planets within the habitable zone, many of these newly discovered
transiting M-dwarf planetary systems will contain additional planets not
seen in transit but are potentially detectable with radial velocity (RV) follow-up
observations. It is therefore of interest to observers conducting such
observations of these systems to quantify the effort which is required to
detect these additional `non-transiting' planets and obtain accurate measurements
of all planet masses.

Unfortunately, RV observations are often deterred by strong 
RV jitter signals \citep{wright05}
associated with magnetic regions which, for M-dwarfs,
tend to be strongest on stars undergoing rapid rotation 
\citep{mohanty03, browning10, reiners12a, west15}\footnote{Alternatively, one
  may consider the correlation between magnetic activity in cool stars with
  the star's Rossby number (a measure of the effect of rotation on convective flows)
  instead of its rotation period \citep[e.g.][]{noyes84}.}. 
For M-dwarfs, empirically there exist two populations of rotation periods 
\citep[e.g.][]{irwin11, mcquillan13a, mcquillan14, newton16a} with 
a significant subset of M-dwarfs having rotation periods $\gtrsim 40$ days. 
These \emph{slowly rotating} 
stars are expected to have low levels of rotationally-induced 
stellar jitter from active regions such as spots and plages \citep{aigrain12} 
compared to their rapidly rotating counterparts.
For example, the slowly rotating M-dwarfs GI 176 and GI 674 
($P_{\mathrm{rot}} = 39$ and 35 days respectively) have been shown to
exhibit comparatively small levels of activity jitter with an amplitude 
of $\sim 5$ \mps{} \citep{bonfils07, forveille09}. 
\emph{The typical low-amplitude jitter thus makes slow rotators 
attractive targets for the atmospheric characterization of their  
planetary companions and detection of additional planets not seen in
transit}, both of which are strongly deterred by stellar jitter. 
Rapid rotators are made additionally disfavorable for RV measurements as the 
intrinsic Doppler broadening of spectral features significantly degrades 
the RV accuracy thus making the determination of planetary masses less precise.

In this study we investigate the observational effort
required to detect additional planets in transiting M-dwarf planetary systems,
focusing on the subset of M-dwarfs which are slowly rotating and are
therefore less likely to be magnetically active. We address this problem
using the GJ 1132 planetary system as a fiducial test case. We do so by
modelling the observed photometric variability of GJ 1132 in order to 
model the corresponding RV jitter from rotationally modulated active regions.
We then perform a suite of Monte-Carlo realizations of simulated RV timeseries 
which sample from 
the known population of small planets around M-dwarfs and compute the detection 
completeness of those planets under realistic observing conditions. 
A direct consequence of this calculation provides 
\emph{limits on the number of RV  
observations required to recover additional planets around slowly rotating M-dwarfs 
such as GJ 1132}. 

In Sect.~\ref{gj1132system} we briefly summarize the current state of knowledge 
of GJ 1132 and GJ 1132b. In Sect.~\ref{mcsims} 
we present the details of our Monte-Carlo simulations, Sect.~\ref{results} presents 
the resulting detection completeness of small planets around GJ 1132 followed in 
Sect.~\ref{discussion} by a discussion of our assumptions and the broad applicability 
of our results to similar planetary 
systems such as those which will be discovered with \emph{TESS}. 
A summary of our results is presented in Sect.~\ref{conclusion} for ease of reference.

\section{GJ 1132 Planetary System} \label{gj1132system}
The GJ 1132 planetary system 
\citep[M4.5V, $V=13.49$, $J=9.25$;][hereafter \citetalias{berta15}]{berta15} 
is the most recent planet discovery 
from the \emph{MEarth} transiting planet survey \citep{irwin15}. At just 12 pc, the 
planet known as GJ 1132b is currently one of the closest transiting rocky exoplanets 
to the solar system. GJ 1132b is slightly larger 
than the Earth and has an orbital period of less than 2 
days, placing it interior to its host star's habitable zone. Photometric and RV  
measurements determined that the planet has a rocky bulk composition 
\citepalias{berta15}. Together, this has 
lead to claims that GJ 1132b might represent a Venus analog (low water vapor mass 
fraction as a result of thermal escape) but requires atmospheric 
characterization via spectroscopic follow-up to probe the planet's atmosphere and  
challenge that notion. Owing 
to the planetary system's close proximity, GJ 1132b represents a very appealing 
target for the atmospheric characterization of a rocky exoplanet with \emph{JWST}. 

As discussed in the introduction, this single planetary system should 
host additional planets not seen in transit which 
could potentially exist within the star's habitable zone 
between $\approx 15-39$ days \citep{kopparapu13}. Scrutinous photometric 
monitoring and a modest RV timeseries 
did not reveal the presence of any additional planets (\citetalias{berta15} 
and this work. See Sect.~\ref{harpsevidence} and Table~\ref{bayesfactors}). 
Although any planets 
co-planar with GJ 1132b, but with $a/R_s \gtrsim 42$, would not transit. 
For reference, $a/R_s=16$ for GJ 1132b.

The planet discovery paper presented 
multiple transit observations and measured a planetary radius of GJ 1132b of 
1.16 R$_{\oplus}$. RV measurements taken with the HARPS spectrograph 
\citep{mayor03} also revealed a 
$3\sigma$ mass detection of 1.62 M$_{\oplus}$ thus giving GJ 1132b a bulk density consistent 
with a rocky, Earth-like composition. Additionally, \citetalias{berta15} presented a long 
baseline stellar light curve for which a stellar rotation period of $\sim 125$ days was 
inferred. From this slow rotation period one expects a correspondingly low level of 
stellar jitter \citep{aigrain12} 
which is greatly beneficial for the detection of additional planetary 
companions with RVs. The properties of GJ 1132 and GJ 1132b are summarized in 
Table~\ref{gj1132table}. 

\begin{deluxetable*}{lcc}
\tabletypesize{\scriptsize}
\tablecaption{GJ 1132 Planetary System Properties\label{gj1132table}}
\tablewidth{0pt}
\tablehead{\textbf{Parameter} & \citetalias{berta15} \textbf{Value} & \textbf{Value from this work}}
\startdata
\textbf{GJ 1132 (star)} & & \\
Stellar Mass, $M_s$ & $0.181 \pm 0.019$ M$_{\odot}$ & - \\ 
Stellar Radius, $R_s$ & $0.207 \pm 0.016$ R$_{\odot}$ & - \\
Effective Temperature, $T_{\mathrm{eff}}$  & $3270 \pm 140$ K & - \\
Stellar Luminosity, $L_{s}$ & $(4.38 \pm 0.34) \times 10^{-3}$ L$_{\odot}$ & - \\
Stellar Age & $>5$ Gyrs & - \\
Rotation Period, $P_{\mathrm{rot}}$ & 125 days & $122.31^{+6.03}_{-5.04}$ days\tablenotemark{a} \\
\tableline
\textbf{GJ 1132b (planet)} & & \\
Orbital Period, $P_b$ & $1.62893 \pm 3.1 \times 10^{-5}$ days & - \\
Time of Mid-Transit, $T0_b$ & $2457184.55786 \pm 0.00032$ BJD & - \\ 
Planetary Radius, $r_{p,b}$ & $1.16 \pm 0.11$ R$_{\oplus}$ & - \\
Eccentricity, $e_b$ & 0 & - \\
Inclination, $i_b$ & $88.63 \pm 0.86^{\circ}$ & - \\
Semiamplitude, $K_b$ & $2.76 \pm 0.92$ \mps{} & $2.69^{+0.81}_{-0.79}$ \mps{} \tablenotemark{a} \\
Planetary Mass, $m_{p,b}$ & $1.62 \pm 0.55$ M$_{\oplus}$ & $1.58^{+0.49}_{-0.48}$ M$_{\oplus}$ \tablenotemark{b}
\enddata
\tablenotetext{a}{Uncertainties are based on the 16$^{\mathrm{th}}$ and 
84$^{\mathrm{th}}$ percentiles of the parameter's marginalized posterior 
distribution.}
\tablenotetext{b}{Uncertainties are propagated from uncertainties in K$_b$, P$_b$, 
M$_s$, and i$_b$.}
\end{deluxetable*}

\section{Monte-Carlo Simulations} \label{mcsims}
To determine the detection completeness of small ($\leq 4$ R$_{\oplus}$)  
planets around GJ 1132 with dedicated RV follow-up, here we run 
an extensive Monte-Carlo (MC) simulation of planetary systems (including GJ 
1132b) around GJ 1132. When constructing RV timeseries and searching 
for additional planets we fully exploit the available data for this system described 
in Sect.~\ref{gj1132system} which includes a model of the RV jitter derived from 
the star's photometric light curve. 
The setup of our MC simulation is described in the proceeding sections.

\subsection{Modelling Radial Velocity Jitter from MEarth Photometry} \label{jitter}
\subsubsection{Gaussian process photometry model}
\begin{figure*}
\centering
\includegraphics[scale=.55]{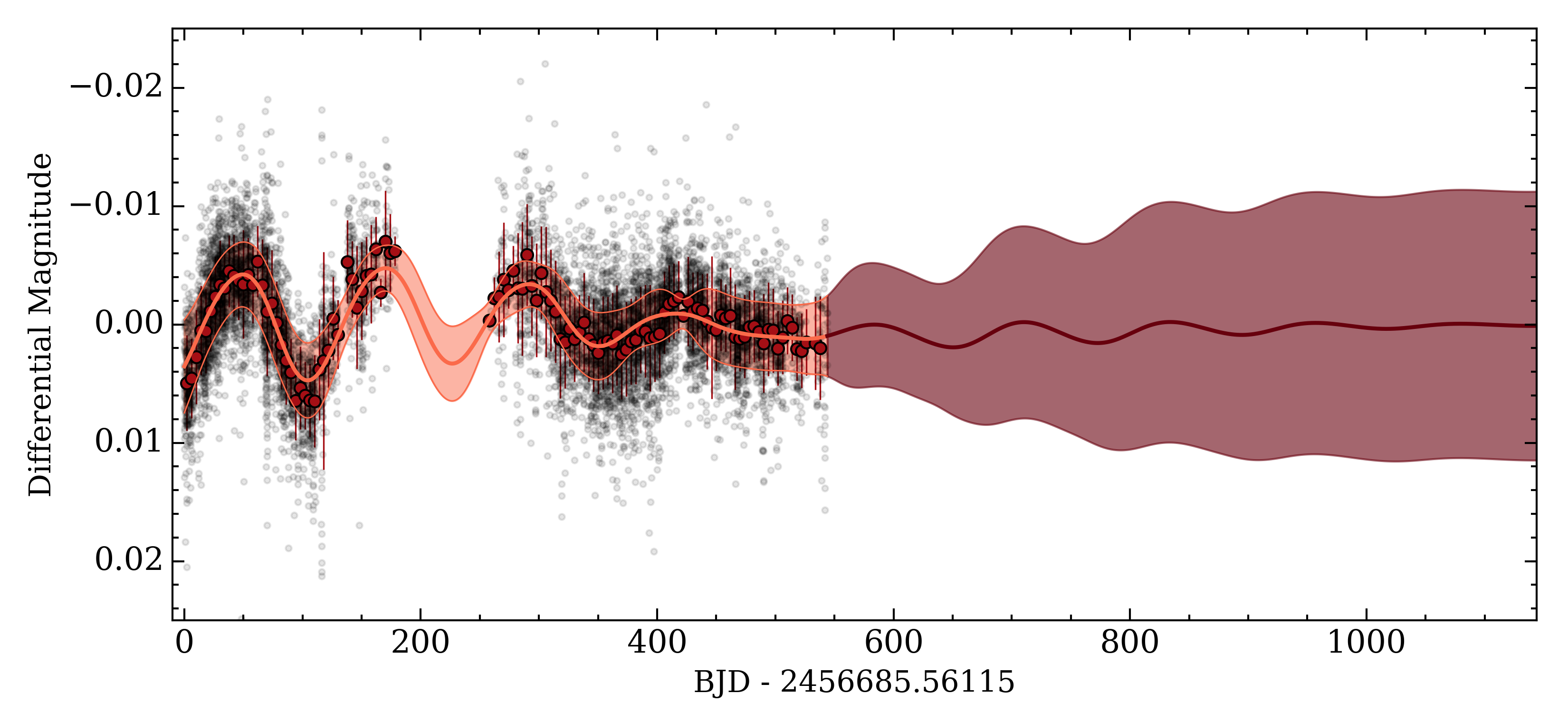}
\caption{The MEarth photometry originally presented in \citetalias{berta15}. 
Red dots are the binned photometric measurements modelled with a 
quasi-periodic Gaussian process (GP) which reveals a stellar rotation period of 
$\sim 122$ days. The mean GP model and 99\% confidence 
intervals are shown in pink. The GP predictive light curve and 99\% 
confidence intervals are shown for 600 days following the last photometric 
observation in red.}
\label{fig:mearthphoto}
\end{figure*}

The rotation of active regions such as spots and plages in the stellar photosphere 
of GJ 1132 gives rise to RV jitter which varies with time.
The long-baseline photometric light 
curve presented in \citetalias{berta15} (planetary transits are removed)
provides the opportunity to model this source of astrophysical noise empirically. 
In brief, this is done by modelling the photometric light 
curve with a smoothly varying function which is then used to predict the associated 
RV signal from the active regions presumed responsible for the observed photometric 
variability. Because the transits of GJ 1132b have
  been removed from the light curve, the observed photometric variability can be solely
  attributed to active regions on the star. Therefore unlike the observed RVs which
  contain signals from GJ 1132b and possibly from additional, non-transiting planets,
  the light curve traces the source of stellar jitter only. 
The light curve and best-fit model, shown in Fig.~\ref{fig:mearthphoto},
are described in the proceeding paragraphs.

The photometric data of GJ 1132 were obtained with the \emph{MEarth}-South telescope 
array at the Cerro Tololo Inter-American Observatory (CTIO) in Chile between 
January 2014 and July 2015. The systematic effects unique to the \emph{MEarth} telescopes 
have been corrected for and are described in detail in \cite{newton16a}. This includes 
the removal of the `common-mode' systematic effect resulting from atmospheric 
extinction due to telluric water vapor absorption in the \emph{MEarth} bandpass. 
We proceed using the fully 
reduced photometry following a clip of `bad' photometric points corresponding 
to frames with an uncharacteristically low zero-point magnitude correction possibly  
resulting from cloud cover.

Due to complexities regarding the differential rotation of the stellar photosphere 
and variations in the sizes and contrast of multiple active regions on the stellar 
surface, the photometric variability need not be strictly periodic. 
We note that for mid M-dwarfs such as GJ 1132, the effect of differential rotation 
may vanish or at least be small due to the extensive convective depth 
\citep{barnes05}. Additionally, parametric 
models of stellar variability due to active regions feature degenerate model 
parameters including the sizes and spatial distribution of active regions 
thus making it difficult to 
accurately constrain model parameters of active regions. We instead opt to model 
the photometric variability of GJ 1132 using non-parametric Gaussian process 
(GP) regression. A discussion of the full Gaussian process formalism is presented in 
appendix~\ref{appendixGP} for the interested reader. 
Most crucially, by assuming that photometric measurements 
are correlated in time via the rotation of the star, we model the covariance between 
photometric measurements with a GP specified by a null mean function 
and a covariance function which varies quasi-periodically (Eq.~\ref{cov}). 

The photometry shown in Fig.~\ref{fig:mearthphoto} is accompanied by the mean
of the GP predictive distribution and its $99$\% confidence interval. We then use 
the GP predictive distribution to predict the photometric variability for 600 days 
following the last available photometric measurement. 

\subsubsection{Radial velocity jitter model} \label{sect:rvjm}
From the photometry, the corresponding RV signal 
or astrophysical jitter can be approximately estimated.
\cite{haywood16} used contemporaneous RV measurements and images of the Sun
to argue that photometry is an incomplete diagnostic of the RV jitter from active
regions. Yet the photometric light curve still gives approximate 
empirical insight into the nature of the star's RV jitter arising
from active regions. 
The conversion to RV from photometry is done using the analytical 
$FF'$ method from \cite{aigrain12} which uses the time-varying 
fractional coverage of the stellar disk by active regions and it first time derivative 
to model the corresponding RV variations. Due to the smoothness of the 
GP model of the light curve we are able to compute the time derivative 
of the star's fractional spot coverage in a simplistic way using finite differences. 
Specifically, we 
use the python \texttt{numpy.gradient} function;
an accurate second-order forward/backwards scheme to estimate the time 
derivative at the boundaries and a central differences scheme elsewhere. 

This jitter model does not consider certain higher order effects such as 
granulation and stellar oscillation modes whose amplitudes are each 
seen to diminish with later spectral types 
\citep{christensen04, gray09, dumusque11} and we assume can be averaged down 
using long integration times \citep{lovis05}. 
Furthermore, we do not include the jitter arising from the Zeeman 
broadening of spectral features in the presence of magnetic fields \citep{reiners13} 
because empirical correlations between the stellar rotation period and magnetic field 
strength indicate that slow rotators, such as GJ 1132, have a negligible contribution to 
the RV jitter from Zeeman broadening \citep{reiners07, west15} 
even at near-IR wavelengths where the effect is stronger than in the optical 
\citep{reiners13}. We also neglect flares in our jitter model because
  their distinctive spectral signature allows them to be easily flagged and removed from
  the RV timeseries \citep{schmidt12}.

The modelled contributions to the RV jitter from the $FF'$ method result from the 
superposition of two effects: i) the suppression of particularly Doppler-shifted 
flux from the occultation of the 
differentially Doppler-shifted stellar limbs by active regions traversing the stellar 
disk (the flux effect) and ii) from the suppression of the convective blueshift at the 
photospheric boundary as the fractional spot coverage evolves with time. 
Focusing on the latter effect, the 
convective blueshift term scales directly with the velocity difference between the 
active regions and the quiet 
photosphere $\delta V_c$, and the ratio of the magnetized area 
to the spot surface $\kappa$. Because we do not have direct observations of the 
spottedness of M-dwarfs \citep{oneal05}, we fix $\delta V_c$ to the solar value of 
$\sim 300$ \mps{} and use the rms about the keplarian model to the RV 
measurements presented in \citetalias{berta15} ($\sim 2.9$ \mps{)} to estimate 
a fiducial value of $\kappa \sim 6$. 
We consider the adopted model parameters to be 
conservative given that the spot-to-photosphere contrast should be smaller 
at near-IR wavelengths compared to the optical and
where upcoming near-IR velocimeters will operate
\citep{martin06, huelamo08, prato08, reiners10, mahmud11}.

\begin{figure*}
\centering
\includegraphics[scale=.55]{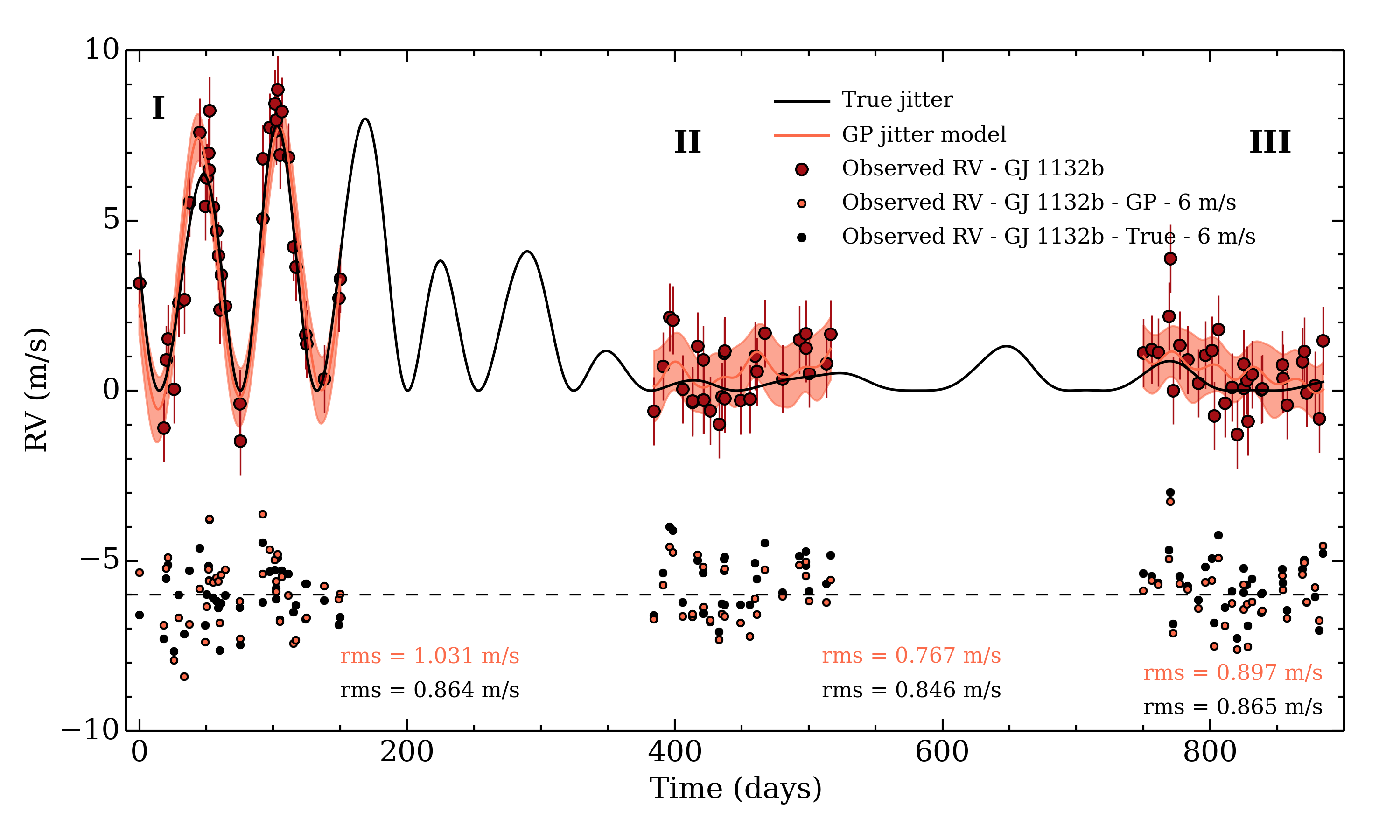}
\caption{An example of a 1-planet simulated RV timeseries on an 
arbitrary time axis and the 
keplarian contribution (from GJ 1132b) removed. The RV measurements (red 
circles) therefore contain stellar jitter, predominantly from the suppression of 
convective blueshift, and noise. The true jitter from which the data are 
derived is shown in black. The optimized quasi-periodic GP jitter model and 95\% 
confidence intervals for each observing window (I,II,III) are shown in pink. 
The residuals about the true jitter are plotted as black points while the 
residuals about the 
GP jitter model are plotted as pink points. Both sets of residuals are shifted
downwards by 6 \mps{} for clarity. The rms of each set of residuals is 
annotated in the panel (residuals about the true jitter and GP jitter model in black and 
pink respectively).} \label{fig:gpjitter}
\end{figure*}

Due to the slow stellar rotation of GJ 1132, the dominant source of 
jitter from active regions results from the suppression of convective blueshift 
and reaches a 
maximum level of $\sim 8$ \mps{} when the star's fractional coverage by active 
regions is at its largest observed value. 
Fig.~\ref{fig:gpjitter} shows our RV jitter model for an example 
RV timeseries. The jitter, if left unfiltered, 
can be a few times greater than the semiamplitude of GJ 1132b.
We would therefore like to model the RV jitter contribution to help
  mitigate its contribution to our simulated RV timeseries and thus allow
  us to detect additional planets. In practise, the functional form of the
  RV jitter is unknown (although we know it in our model because it was derived from
  the GP photometry model in Fig.~\ref{fig:mearthphoto}) so we opt to use a second
  quasi-periodic GP, trained on the photometry, to model the RV residuals
  after the removal of the GJ 1132b keplarian model (i.e. the GP mean function 
  is the GJ 1132b keplarian model). We'll refer to this second GP as the
  \emph{GP jitter model} throughout.
This methodology has been successfully demonstrated in 
the literature on stars more active than GJ 1132 
\citep[e.g.][]{baluev13, haywood14, grunblatt15}. 

In Fig.~\ref{fig:gpjitter} we distinguish adjacent observing seasons
by labelling them I, II, and III respectively. This is necessary because the
GP jitter model behaves differently depending on 
the amplitude of the RV variation compared to the measurement 
uncertainties ($\sigma_{\mathrm{RV}}=1$ \mps{} in Fig.~\ref{fig:gpjitter}) and the
jitter amplitude itself varies between the three seasons. 
Fig.~\ref{fig:mearthphoto} revealed a long decaying timescale of the fractional 
coverage by active regions implying that the RV jitter will vary 
between stellar rotation cycles. 
In regions where the RV measurement amplitude is small (labelled II and III),
the measurement dispersion is comparable to the underlying true jitter
(\emph{black line}) which causes the GP jitter model (\emph{pink line})
to (incorrectly) find high-frequency structure that is not present in the 
true jitter; GP over-fitting. However, because the amplitude 
of the best-fit GP jitter model is comparable to the true jitter, the measurement 
residuals about the GP jitter model (\emph{pink dots}) and the true jitter 
(\emph{black points}) are consistent
and on the order of $\sigma_{\mathrm{RV}}$ ($\sim 0.9$ \mps{)}.

When the star is largely covered in active regions (labelled 
  I), the jitter amplitude is $> \sigma_{\mathrm{RV}}$ and a GP is useful for 
  modelling the jitter. In this case the residual rms about the GP jitter model
  is reduced to the 
level of $\sim 1$ \mps{} and crucially, comparable to the residual rms 
of the observed RVs about the true jitter in the `quiet' regions. 
\emph{Therefore we confirm that the use of a GP in modelling the jitter is an 
effective method of mitigating the jitter down to the level of the RV 
measurement uncertainties.} In Sect.~\ref{sect:rapid} we elaborate on the 
power of GP jitter models applied to rapid rotators where the jitter amplitude 
is significantly larger than it is for the slowly rotating GJ 1132.

\subsection{Radial Velocity Timeseries Construction} 
\subsubsection{Radial velocity contributions}
In order to simulate a suite of planetary systems that could potentially exist in  
the GJ 1132 system, we construct a sample of hypothetical planetary systems and 
simulate their RV signals with jitter included. 
We refer to these components as the astrophysical terms in the timeseries.
The astrophysical terms therefore include the keplarian signal from 
GJ 1132b, the aforementioned additional keplarian signals from 
hypothetical planets 
(see Sect.~\ref{gj1132cs} for a description of these planets), and the 
stellar RV jitter due to rotating active regions derived in 
Sect.~\ref{jitter}. Each astrophysical 
term is assumed to be independent. 

In addition to these astrophysical sources, we add a 
noise model containing a white noise term with a fixed variance of 1 \mps{} 
and a systematic noise term 
which permits us to vary the global noise properties of each timeseries in a simple, 
parameterized way. 
In practice, the total non-astrophysical  noise injected into each 
timeseries varies from $\sigma_{\mathrm{RV}} \in [1,2]$ \mps{.} 
Given the late spectral type of GJ 1132b, a 
RV precision of 1 \mps{} should be achievable \citep{figueira16}. 
The variance in the Gaussian noise term is chosen to resemble 
the long-term stability limit of current state-of-the-art velocimeters 
\citep[e.g. HARPS;][]{mayor03} 
which is also equal to the anticipated stability limit of some upcoming near-IR 
velocimeters 
(e.g. \spirou{;} \citealt{delfosse13, artigau14}, \emph{Habitable Zone Planet Finder}; 
\cite{mahadevan12}, \emph{Infrared Doppler instrument}; 
\cite{kotani14}, \emph{CARMENES}; \citealt{quirrenbach14}).

\subsubsection{Time-sampling}
For each simulated planetary system we observe the star \nobs{} times 
with \nobs{} $\in [40,300]$. The adopted limits on \nobs{} are motivated by 
the approximate number of RV measurements required to detect a planet in 
a blind survey and an assumed upper limit on a reasonable number of RV measurements
per star respectively. A modest limit on the maximum value of \nobs{} also limits the 
computational cost of our study. 
The observations are taken at most twice per night over the 
course of 3 years, similar to the time baseline of a dedicated exoplanet 
RV survey. The observation timestamps are randomly drawn from the 
window function (time sampling) generated from the ephemeris of GJ 1132. 
Although the window function is based on the 
location of the HARPS-South instrument at La Silla 
Observatory in Chile, we exclude observation epochs during dark-time due to the 
priority reserved for imaging instruments during that time. Although this restriction 
does not affect the La Silla site, it will be applicable to instruments such as \spirou{} 
at \emph{CFHT}. From preliminary tests which relaxed the exclusion of dark-time 
epochs, this restriction was found to not have a significant effect 
on our results. 

\subsection{Planet Sample} \label{gj1132cs}
In order to investigate the potential detection of additional planets around GJ 
1132, we construct a sample of small planets 
according to the occurrence rates from the primary Kepler mission 
\citep{dressing15a}. We opt to use the planet occurrence rates derived 
from the Kepler transit 
survey, coupled to a planetary mass-radius relationship, rather than the occurrence rates 
derived from radial velocity surveys \citep[e.g.][]{bonfils13}. This selection 
owes itself to the improved statistics resulting from the larger sample of transiting 
planets than radial velocity planets around M-dwarfs. 

We find that dynamical stability arguments require 
that the planetary orbital periods $P$ of our simulated planets always
be greater than that of GJ 1132b. 
Adopting the grid of planet radii $r_p$ and orbital periods 
from \cite{dressing15a} restricts our analysis to $r_p \in [0.5,4]$ R$_{\oplus}$ 
and $P \in (P_{\mathrm{b}}, 200]$ days where the subscript $b$ refers to GJ 1132b. 
To derive the expected RV signal from each simulated planet we convert 
$r_p$ to planet mass $m_p$ by simply assuming 
an Earth-like bulk density for planets with $r_p \leq 1.6$ R$_{\oplus}$. 
The highest precision exoplanet mass and radius measurements ($\leq 20$\% 
uncertainty) have shown that the upper limit on $r_p$ for rocky worlds (little-to-no 
volatiles) is $\sim 1.6$ R$_{\oplus}$, 
below which most exoplanets have an Earth-like bulk density \citep[$\sim 17$\% 
Fe and 83\% MgSiO$_3$ mass fraction;][]{zeng13, charbonneau15, dressing15b}. This 
boundary has also been approximately confirmed from theoretical studies 
\citep[e.g.][]{lopez14}. The 
subset of planets with $r_p > 1.6$ R$_{\oplus}$ are assigned the bulk density of 
Neptune; a solid core surrounded by a gaseous envelop rich in volatiles.  
We refer to these two classes of planets as Earth-like 
($\bar{\rho} = 5.51$ g cm$^{-3}$) and Neptune-like ($\bar{\rho}=1.64$ 
g cm$^{-3}$) compositions respectively. 
Adopting this mass-radius relationship represents a conservative choice in the sense 
that if instead we adopt a probabilistic mass-radius relationship motivated by the 
empirical distribution of exoplanetary masses and radii, such as that from 
\cite{weiss14}, 
the resulting planetary masses are often found to be greater than those which are 
obtained when assuming either an Earth-like or Neptune-like bulk density. 
Therefore, for a fixed  
$r_p$ the planet is more easily detectable assuming the mass-radius relationship 
from \cite{weiss14} thus making the adopted mass-radius relationship a conservative 
choice. The implications of changing the mass-radius relationship are discussed in 
detail in Sect.~\ref{mrrel}.

\subsubsection{Dynamical considerations}
At the orbital periods considered, the circularization timescale of simulated planets 
need not be less than the assumed age of the system ($\sim 5$ Gyrs; \citetalias{berta15}).
Therefore we do not fix simulated planets on circular orbits and instead randomly draw  
orbital eccentricities from the Beta distribution in \cite{cloutier15} used to 
describe the eccentricity distribution of RV exoplanets detected with 
high significance \citep{kipping13}. 
Assuming planets around GJ 1132 are remnant bodies from a primordial debris disk 
that is collisionally relaxed following the planet formation process 
\citep{raymond05, kokubo06}, the eccentricity dispersion within any planetary 
system $\langle e^2 \rangle$ is related to the dispersion in orbital inclinations 
via $\langle i^2 \rangle \sim \langle e^2 \rangle /4$ radians 
\citep{stewart00, quillen07} 
which we use to sample the mutual inclinations of drawn planets. Furthermore, because 
small exoplanets around M-dwarfs in multiple systems have been shown to have lower  
eccentricities than their giant planet counterparts \citep{kane12, vaneylen15}, 
we focus on orbital eccentricities $\lesssim 0.5$. 
This also ensures that the inclinations of simulated planetary systems resemble 
the empirical distribution containing mostly low mutually inclined compact systems 
\citep{figueira12, fabrycky14}.

Because in our test case 
we are interested in detecting additional planets in a planetary system 
with one confirmed planet, dynamical stability arguments further restrict the 
types of multi-planet systems that can potentially be present around GJ 1132. 
We therefore impose additional, analytical  
constraints on simulated planetary systems in an attempt to ensure that simulated 
planetary systems are dynamically stable without having to perform detailed 
N-body simulations of every simulated system. 
Specifically, we constrain the simulated planet eccentricities 
to ensure that the orbits of GJ 1132b and the added planets do not cross. We also 
insist that each planetary system be Lagrange stable which 
naturally implies Hill stability \citep{marchal82, duncan89, gladman93}.  
Lagrange stability requires that the ordering of planets remains fixed, that both 
planets remain bound to the central star, and also places limits on 
permissible changes in planets' semimajor axes and eccentricities making 
the Lagrange stability criterion a more stringent definition of stability than 
Hill stability alone. 
The approximate condition for Lagrange stability is 
$\delta \gtrsim 1.1 \delta_{\mathrm{crit}}$ \citep{barnes06} where 
$\delta=\sqrt{a_{\mathrm{out}}/a_{\mathrm{in}}}$ and $\delta_{\mathrm{crit}}$ is the 
value of $\delta$ which satisfies

\begin{equation}
\alpha^{-3}(\mu_1 + \frac{\mu_2}{\delta_{\mathrm{crit}}^2}) 
(\mu_1 \gamma_1 + \mu_2 \gamma_2\delta_{\mathrm{crit}}^2) 
= 1+ \mu_1 \mu_2 \left( \frac{3}{\alpha} \right)^{4/3},
\end{equation}

\noindent where $\mu_i = m_{p,i} /M_s$, 
$\alpha = \mu_1+\mu_2$, and $\gamma_i = \sqrt{1-e_i^2}$ \citep{gladman93}.
The Lagrange stability criterion is therefore uniquely determined by the system body 
masses, orbital separations, and eccentricities. 
Strictly speaking, the analytical Lagrange 
stability criterion is only applicable to three-body systems (star + two planets). 
We therefore apply the stability criterion to adjacent pairs of planets in 
systems with multiplicity $> 2$ noting however that such systems may include 
significant planet-planet interactions whose effects on system stability are not 
captured by the analytic Lagrange stability criterion. 

As there is no analytical stability criterion for systems with multiplicity $> 2$, 
we supplement the Lagrange stability criterion with the heuristic 
stability criterion from \cite{fabrycky12} which is derived from suites of 
numerical integrations of high multiplicity systems \citep{smith09}. 
\cite{fabrycky12} 
reported that a system with $\Delta_{\mathrm{in}}+\Delta_{\mathrm{out}}>18$ will be 
dynamically stable where $\Delta = (a_{\mathrm{out}}-a_{\mathrm{in}})/R_{\mathrm{Hill}}$, 
the subscripts `in' and `out' refer to the inner and outer planet respectively, and 
$R_{\mathrm{Hill}}$ is the mutual Hill radius of adjacent planet pairs. 
Applying these conditions to the ensemble of simulated planetary 
systems results in a planet sample that is weakly biased away from short-period, 
massive planets. The resulting planet distribution is therefore not equivalent to 
the initial planet distribution of \cite{dressing15a}.

All together, these limiting conditions help to ensure that additional planets 
around GJ 1132 do not invoke a strong 
gravitational interaction on GJ 1132b. This is motivated by the fact that 
the GJ 1132b transit light curves do not exhibit any significant transit 
timing or transit duration variations \citepalias{berta15}. 
This constraint greatly simplifies the analysis as it implies that the total 
RV signal resulting from multiple planetary companions  
can be sufficiently described by the superposition of keplarian orbits rather 
than a computationally expensive, fully dynamical model. To facilitate this 
methodology of constructing synthetic RV timeseries via multiple 
keplarian orbits we also demand that planet pairs not exist close to any 
low-order mean-motion resonance wherein the keplarian approximation would 
break down as evidenced by the large associated transit-timing variation 
amplitudes in low mutually inclined systems \citep{agol05, payne10}.

\subsubsection{Planet sample at a glance}
\begin{figure}
\centering
\includegraphics[scale=.55]{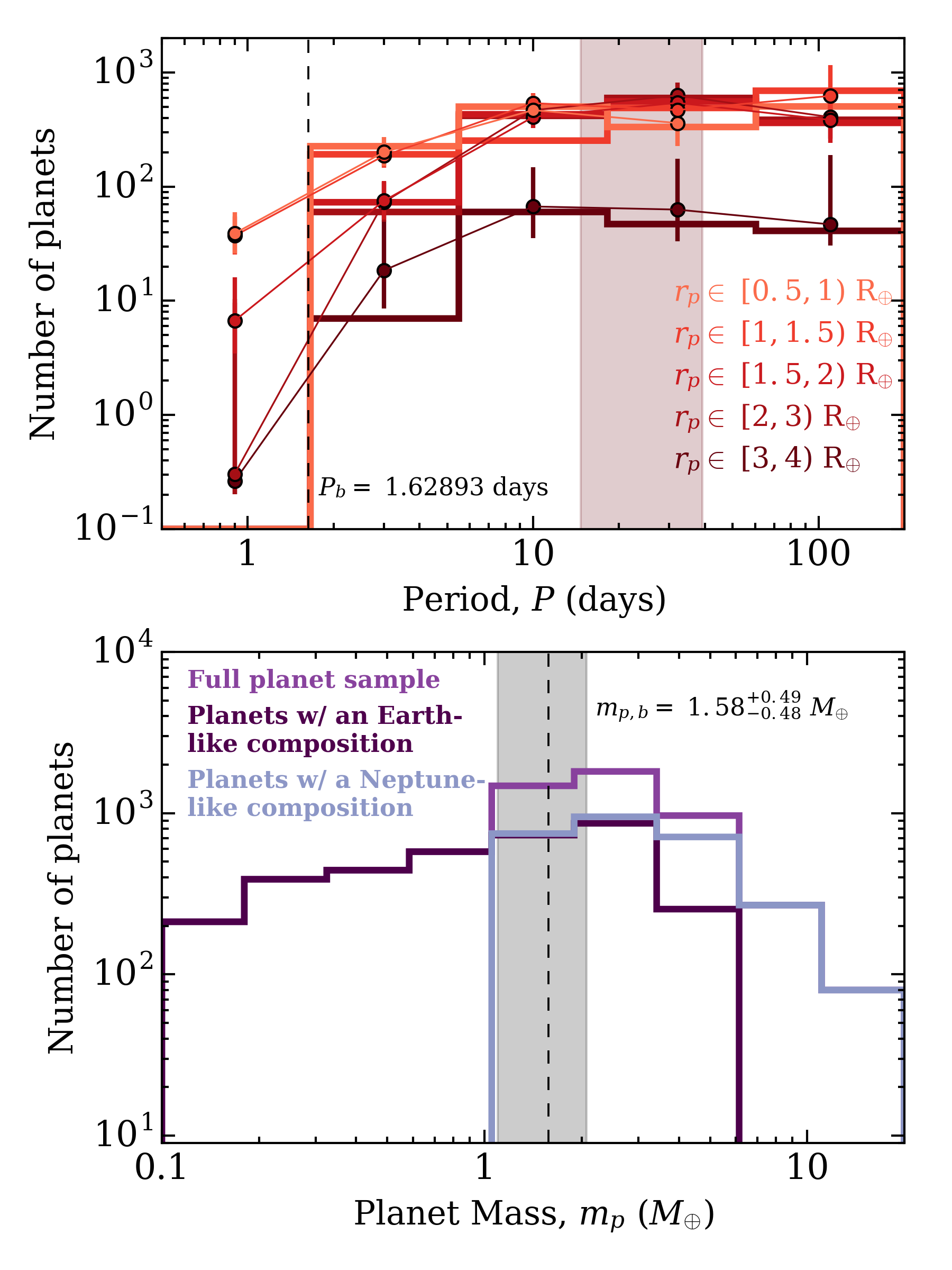}
\caption{Distributions of simulated planet properties. \emph{Top}: orbital 
periods in various planet radius bins. The circular markers represent the 
scaled period distribution from \cite{dressing15a} in identical radius bins. 
The vertical dashed line depicts the orbital period of GJ 1132b. 
The shaded region highlights 
the adopted definition of the star's habitable zone (see Sect.~\ref{hz}; 
$14.7-39.2$ days) \emph{Bottom}: the full sample of planet 
masses along with the sub-samples of rocky and gaseous 
planets. The vertical dashed line and shaded region depicts the 
MAP value and $1\sigma$ uncertainty of the GJ 1132b planet mass from our 
Gaussian process + keplarian fit. (Sect.~\ref{pdetect}).}
\label{planetdist}
\end{figure}

The full distributions of simulated planet orbital periods and masses are shown in 
Fig.~\ref{planetdist}. Orbital periods are shown in various planet radius bins 
and are compared to the \cite{dressing15a} distribution from which they were 
initially drawn. 
Unsurprisingly, our sample is consistent with that of \cite{dressing15a} 
within reported uncertainties for periods $\gtrsim 5$ days. For massive planets 
($r_p \geq 3$ R$_{\oplus}$) 
with orbital periods below this limit but still $> P_b$, issues regarding 
dynamical stability between the simulated planet and GJ 1132b begin to limit 
the planet sample there. The same stability argument results in no simulated
planets with orbital periods 
$< P_b$ in the GJ 1132 system despite there non-zero occurrence rate.

Based on the adopted mass-radius relation, the distribution of planet masses
in the simulation 
is split roughly evenly between rocky and volatile-rich planetary 
compositions with a slight excess of rocky planets ($\sim 55$\% rocky). 
Furthermore, the distribution of simulated planet masses has a median 
value of $\sim 1.9$ M$_{\oplus}$ and the subset of the most massive 
planets ($m_p \gtrsim 4$ M$_{\oplus}$), hence the most easily detectable with 
radial velocities, are not expected to have a rocky bulk composition. 

\subsection{Planet Detection} \label{pdetect}
In order to detect additional planets around GJ 1132 in the synthetic RV 
timeseries in our MC simulations, we compute the Bayesian model evidences for models 
with an increasing number of keplarian orbital solutions (i.e. no planets, 1 planet, 
and 2 planets). From these values we can determine the optimal number of planets 
supported by each dataset and thus the number of detected planets. This method 
is commonly used to detect planetary signals in RV 
timeseries \citep[e.g.][]{ford07, feroz14}. 

We reserve a more complete description of the adopted Bayesian model selection formalism 
for appendix~\ref{appendixZ}. To summarize, the Bayesian model evidence $Z_k$ 
(Eq.~\ref{inte}) 
for each independent model $M_k$ where $k=0,1,2$ is the number of keplarian solutions 
in the model, is computed using the \texttt{Multinest} \citep{feroz08, feroz09, feroz14} 
nested sampling algorithm 
to sample from the unique model parameter posteriors and efficiently calculate the 
model evidences from a specified logarithmic likelihood function (Eq.~\ref{loglike}) 
and appropriately chosen model parameter priors. 

For the single planet model $M_1$, the set of model parameters 
$(P, T0, K)$ corresponds to the measured values for GJ 1132b and are 
therefore well constrained by the transit light curves and RV measurements 
from \citetalias{berta15}. For each parameter we adopt a uniform prior within 
the parameter's $1\sigma$ uncertainty from Table~\ref{gj1132table}. 

The case is somewhat more complicated for the 2-planet model. Because the 
expense of calculating the evidence integral in Eq.~\ref{inte} grows rapidly with the 
dimensionality of the model 
and the domain of the model parameter space, we restrict the analysis to 
a subset of parameter values 
obtained from the results of a Lomb-Scargle periodogram 
\citep{scargle82, gilliland87} analysis of the residual RV 
dataset following the removal of 
the GJ 1132b keplarian signal. Namely, we construct a broad uniform prior on 
$P_c$ (we label the second planet in the $M_2$ model with the subscript $c$)
centered on the most significant 
periodogram peak $P_{\mathrm{peak}}$ with a false-alarm probability (FAP) 
$\leq 0.01$. Each peak's FAP is computed using the analytical 
approximation of \citep{baluev08, meschiari09}.  
The analytical estimation for small FAPs (FAP $\lesssim 0.001$) is known to be 
inaccurate \citep{cumming04} so instead we settle for 
an upper bound on the FAP which is still indicative of a period `detection'. 
Limits on the period prior are placed at $\pm 5$\% which is intended to 
span the peak value without being too restrictive as the location of the peak in the 
discrete periodogram need not correspond exactly to the period of the input 
planet. Furthermore, we avoid peaks located within 2\% of the stellar rotation period 
or its low-order harmonics (i.e. 122.31, 61.15, 40.77, 30.58 days). 
We then adopt broad uniform priors of 
$T0_c \in [T0_b-P_{\mathrm{peak}}/2, T0_b+P_{\mathrm{peak}}/2]$ BJD and 
$K_c \in (0, 15]$ m s$^{-1}$. 
In this way, the detection of additional planets around GJ 1132 requires a small 
false-alarm probability peak in the Lomb-Scargle periodogram of the timeseries as 
well as favorable evidence for the two planet model. 

From the calculated model evidences we compute the Bayes factor, or the ratio of 
evidences $R_{mn} = Z_m / Z_n$ (see also Eq.~\ref{Bfact}), for arbitrary models $m,n=0,1,2$. 
Using the interpretation scheme outlined in 
\cite{kass95}, we claim that a hypothetical second planet, not detected in transit, 
is marginally detected 
in a RV dataset if $R_{21} > 3$ and is a bona fide detection if $R_{21} > 20$.

\subsubsection{Planet detection in the published radial velocities of GJ 1132} \label{harpsevidence}
As a check, we apply our planet detection method to the 25 published RV measurements 
from \citetalias{berta15} using a GP, trained on the MEarth photometry, to model the 
correlated components in the RV data arising from 
stellar rotation. The resulting semiamplitude and planet mass of GJ 1132b are consistent 
with the results from \citepalias{berta15} implying that use of a GP noise model does not 
have a significant effect on the recovery of planet parameters for this particular 
system. This is unsurprising given that the RV data were obtained over only $\sim 1/3$ 
of a rotation cycle and the errorbars on those measurements were comparable to the 
expected RV jitter from active regions (Sect.~\ref{sect:rvjm}). 
The maximum \emph{a posteriori} (MAP) values of $K_b$ and $m_{p,b}$ from this analysis are
reported in Table~\ref{gj1132table}. 

As an additional check, 
this analysis also confirmed that a single planet RV model is strongly 
favored over the null hypothesis using the aforementioned priors on the orbital 
period and time of mid-transit from the transit lightcurves 
\citepalias{berta15} to reduce the computation 
time, but expanding the prior on $K_b$ to $\in (0, 15]$ m s$^{-1}$. 
The single planet model also shows positive evidence compared to 
a 2-planet model indicating that a `GJ 1132c' is not detected in the dataset. 
This highlights the fact that the eventual detection of 
additional planets in the GJ 1132 system will require extensive RV 
follow-up observations. The complete model evidences based on the published RV 
data are given in Table~\ref{bayesfactors}. 

\begin{deluxetable*}{cccc}
\tabletypesize{\scriptsize}
\tablecaption{Model evidences $Z$ based on the published GJ 1132 radial velocity 
timeseries\label{bayesfactors}}
\tablewidth{0pt}
\tablehead{Model & $\ln{Z}$ & Bayes factor & Interpretation}
\startdata
No planets ($M_0$) & $-48.06 \pm 0.01$ & - & - \\
One planet ($M_1$; GJ 1132b) & $-45.05 \pm 0.02$ & $R_{10} = 20.31 \pm 1.01$ & The 
single planet model is \emph{strongly} \\ 
& & & favoured over the null hypothesis. \\
Two planets ($M_2$) & $-46.48 \pm 0.03$ & $R_{12} = 4.16 \pm 0.99$ & The single 
planet model is \emph{positively} \\
& & & favoured over the two planet model.
\enddata
\end{deluxetable*}

\section{Results} \label{results}
The detection completeness to additional planets in the known 
transiting system is computed using $10^3$ realizations of our MC 
simulation for each of the three values of the RV measurement uncertainty  
$\sigma_{\mathrm{RV}} = 1, 1.5, 2$ \mps{}, which are calculated by the quadrature sum of 
parameterized white and systematic noise sources. Uncertainties on the 
computed detection completeness are calculated using Poisson statistics. 
We find that planetary systems with the minimum multiplicity of one  
\emph{never} find strong evidence for an additional planet. These systems are included 
in the completeness calculation because the null detection of a non-existent planet 
should not decrease the detection completeness. 
For systems with multiplicity $>2$, we only focus on 
detecting one of the simulated planets.
The planet assigned to be the `target' planet is simply 
the planet with the largest semiamplitude and will therefore have the most significant 
contribution to a periodogram of the timeseries. 
We do not attempt to detect more than two 
planets given the large size of the planetary system sample and the increased computation 
required to compute model evidences in systems with multiplicity $>2$. 

\subsection{Full Planet Detection Completeness}
\begin{figure*}
\centering
\includegraphics[scale=.65]{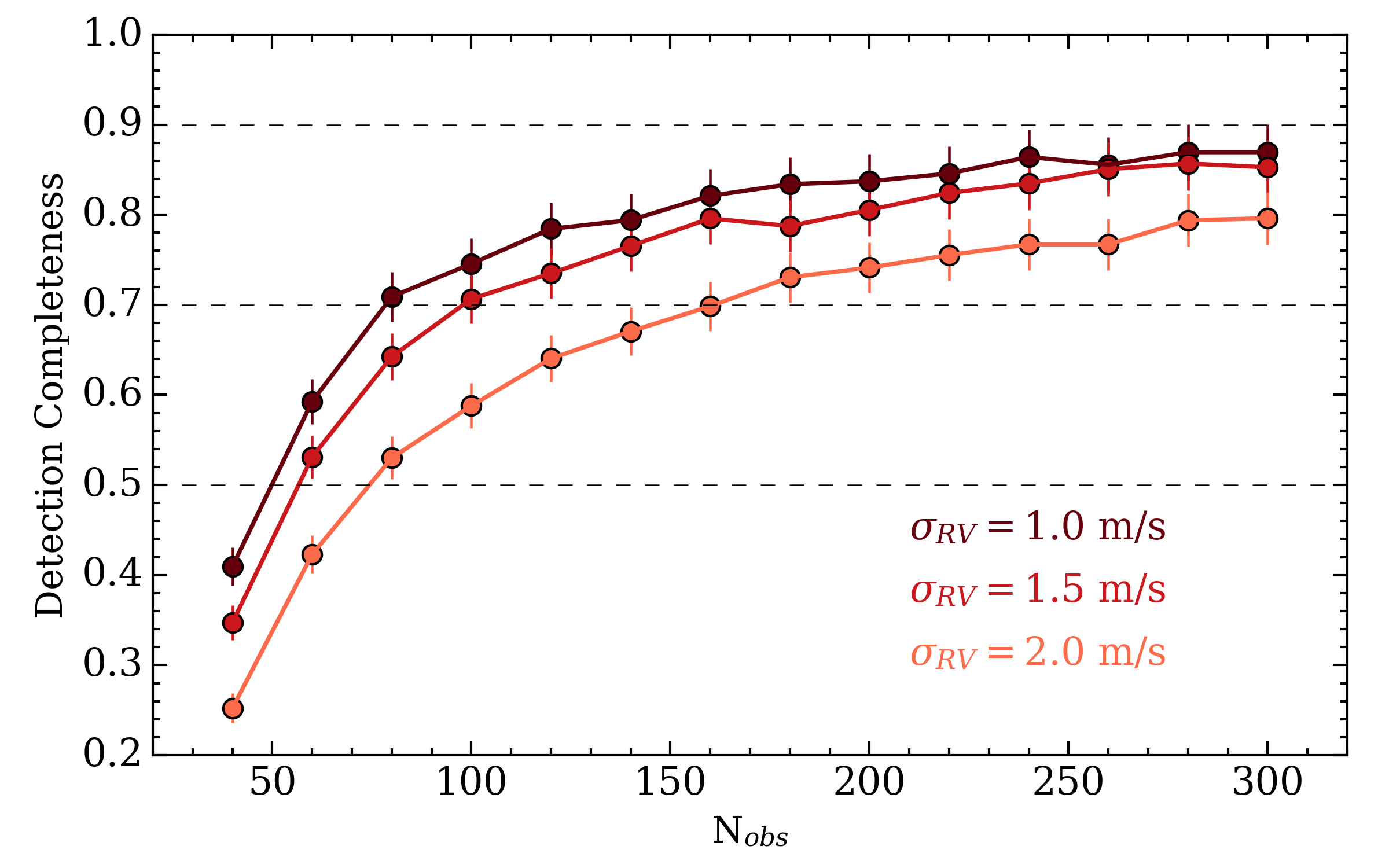}
\caption{The detection completeness for the entire sample ($r_p \in [0.5,4]$ R$_{\oplus}$ 
and $P \in (P_{\mathrm{b}}, 200]$ days) of simulated planets not seen in transit 
around GJ 1132 for various values of the RV measurement uncertainty. Errorbars are 
from Poisson statistics. Dashed 
horizontal lines at 50, 70, and 90 per cent completeness are included to guide the eye.}
\label{detfreq}
\end{figure*}

The detection completeness of our full MC planet sample is shown in 
Fig.~\ref{detfreq}. 
Initially, the detection completeness curves increase 
rapidly with the initial slope of the curve decreasing with $\sigma_{\mathrm{RV}}$ implying 
that as the noise properties improve, one achieves a favorable detection completeness 
with fewer RV measurements. The detection completeness curves asymptotically approach 
maximum values which tend to increase with $\sigma_{\mathrm{RV}}$ as expected. 

\emph{In the best case scenario where the RV measurement uncertainty
is equal to the long-term instrument stability limit of 
$\sigma_{\mathrm{RV}}=1$ \mps{,} we are sensitive to $\geq 50$\% of all additional 
planets even with as little as 50 RV measurements.} In this case 
the maximum detection completeness of $\sim 85$\% is achieved for 
\nobs{} $\geq 200$. 

Similarly, for the largest considered RV measurement uncertainty of 
$\sigma_{\mathrm{RV}}=2$ \mps{,} the detection completeness reaches $\sim 50$\% by 
\nobs{} $\sim 75$ and 
continues to a maximum of 80\% which is $\sim 7$\% less than what is achieved 
with $\sigma_{\mathrm{RV}}=1$ \mps{.} In general, one loses just $\sim 7-15$\% of 
detections across all \nobs{} with an increase in $\sigma_{\mathrm{RV}}$ from 1 to 
2 \mps{.} Depending on the goals of an observer 
regarding their desired detection completeness, the detriment exhibited by increasing 
$\sigma_{\mathrm{RV}}$ to 2 \mps{} may not be so harmful to the results. 
Furthermore, assuming that observational uncertainties are dominated by shot 
noise, settling for $\sigma_{\mathrm{RV}} = 2$ \mps{} only requires $\sim 1/4$ of the 
integration time as for $\sigma_{\mathrm{RV}} =1$ \mps{} \citep{bouchy01} thus allowing 
for a deeper planet detection completeness to be achieved with the same amount of 
telescope time.

\subsection{Detection Completeness in Semiamplitude Bins}
\begin{figure*}
\centering
\includegraphics[scale=.55]{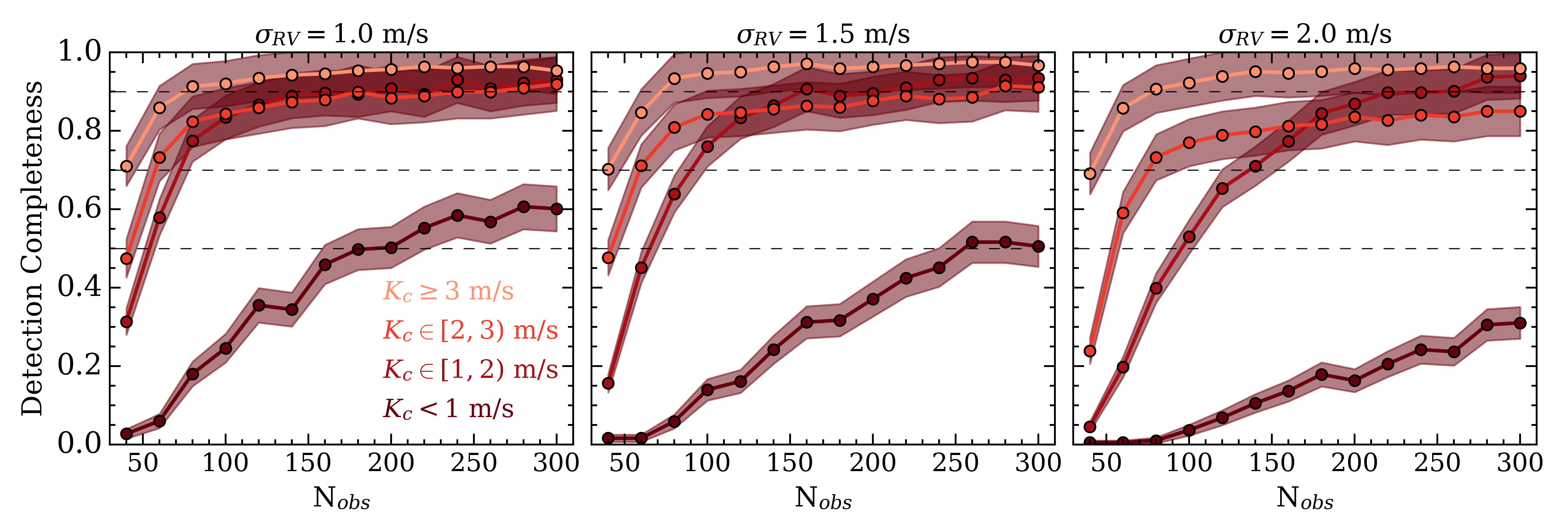}
\caption{Same as Fig.~\ref{detfreq} but for planets in various 
RV semiamplitude bins. Each panel reports the completeness for a
sample with the RV measurement uncertainty annotated 
above the panel. The shaded regions depict the errors from Poisson statistics. 
Dashed horizontal lines at 50, 70, and 90 per cent completeness are 
included to guide the eye.}
\label{detfreqKs}
\end{figure*}

Here we bin our planet sample into four semiamplitude bins each containing 
roughly the same fraction of the full planetary sample ($20-30$\% per bin). 
The resulting detection completeness for each value of $\sigma_{\mathrm{RV}}$ is shown 
in Fig.~\ref{detfreqKs}. Note that the Poisson errors on the curves are larger 
than in Fig.~\ref{detfreq} as a result of the binning.

Planets with $K_c < 1$ \mps{} have signals below the minimum RV measurement 
uncertainty and are hence difficult to detect though not impossible. In the cases 
of $\sigma_{\mathrm{RV}} < 2$ \mps{,} we can 
achieve $> 50$\% detection completeness with \nobs{} $< 300$. 
Planets with $K_c \geq 1$ \mps{} 
represent the planets which are nominally detectable with current instrumentation 
and contribute the most to the detect completeness curves presented in 
Fig.~\ref{detfreq}. As expected, the planet detection completeness to planets with 
larger RV signals increases in general. However, when the level of RV
measurement uncertainty becomes 
comparable to the mean semiamplitude of a bin in Fig.~\ref{detfreqKs} (second and third 
panels), there can be some 
confusion in the completeness curves wherein planets with smaller $K$ appear to have 
improved detection statistics. We note however that these confused curves are still 
consistent within their Poisson errors.

From the completeness curves presented in Fig.~\ref{detfreqKs}, we claim that it 
is not an unreasonable observational task to detect $\gtrsim 80$\% (50\%) 
of planets in a system like GJ 1132 with $K_c \geq 1$ \mps{} for 
$\sigma_{\mathrm{RV}} = 1$ (2) \mps{.}

\subsection{Detection Completeness of Potentially Habitable Planets} \label{hz}
The habitable zone (HZ) around a given star refers to the orbital separations at which 
liquid water can exist on a planet's surface \citep{dole64, hart79}. Yet, given that 
simple definition the boundaries of the HZ have not been unambiguously 
well-defined. This is understandable given the multitude of parameters that can affect 
the planetary surface conditions. Among these effects 
are the atmospheric mass and composition 
which can alter the planetary surface pressure and thus the relevant temperatures for 
phases changes of water \citep{vladilo13}. The presence of clouds can also have profound 
effects on the heating and cooling of the planet and therefore alter the orbital 
separations of the HZ \citep{selsis07, yang13}. 
Certain geometries of the system, 
including planet obliquity \citep{williams97, spiegel08, spiegel09} and orbital 
eccentricity \citep{dressing10, cowan12}, 
may also have an effect on the long term habitability of a planet. 

For this study, we adopt a conservative definition of the HZ based on the parametric 
model of \cite{kopparapu13} which is derived using a 1D radiative-convective climate 
model in the absence of clouds. Following \cite{kasting93}, the inner edge of the HZ is 
defined by the `water-loss' or moist greenhouse limit wherein an increase in insolation 
would result in the photolysis of water in the upper atmosphere and result in subsequent 
hydrogen escape. We note that the inclusion of clouds may decrease the HZ inner edge as 
a result of the reflective properties of clouds. 
The outer edge of the HZ is determined by the maximum greenhouse limit 
wherein an increase in atmospheric CO$_2$ levels results in a net cooling as the 
effect of the increased albedo from Rayleigh scattering dominates over the additional 
greenhouse effect. We performed a small suite of tests using alternative 
parameterizations of the inner HZ edge from \cite{yang14} (for a cloudy, slowly rotating 
planet. These tests 
concluded that our completeness results are not strongly affected by the 
various assumptions regarding the HZ.

\begin{figure*}
\centering
\includegraphics[scale=.55]{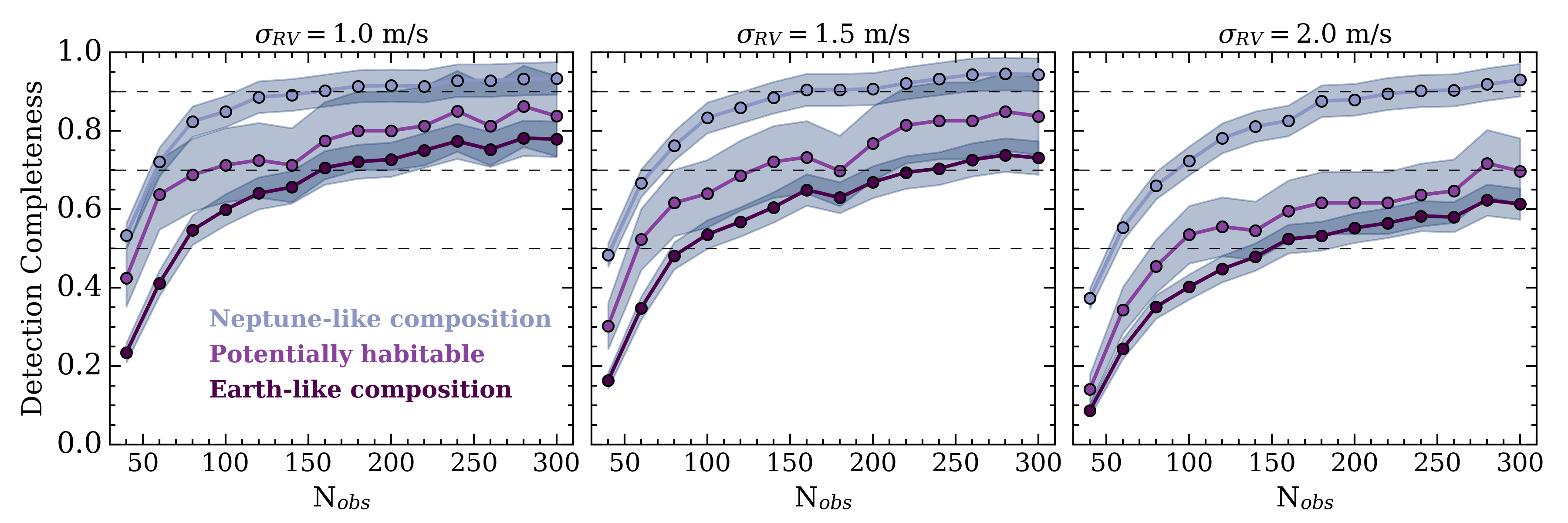}
\caption{Same as Fig.~\ref{detfreq} but for the subset of planets that i) have 
volatile-rich envelops (\emph{Neptune-like composition}), ii) are rocky 
(\emph{Earth-like composition}), and iii) are both 
rocky and lie within the habitable zone of GJ 1132. Each panel reports 
the completeness for a sample with the RV measurement uncertainty annotated above the panel. 
The shaded regions depict the errors from Poisson statistics. 
Dashed horizontal lines at 50, 70, and 90 per cent completeness are included to guide 
the eye.}
\label{detfreqcomp}
\end{figure*}

Assuming for GJ 1132 the stellar effective temperature and luminosity from 
Table~\ref{gj1132table}, the resulting inner and outer HZ limits are 14.7 and 
39.2 days respectively. We compile the set of potentially habitable planets by 
isolating the subset of planets which are rocky and whose orbital periods lie within the 
aforementioned HZ bounds. The detection completeness for this sub-sample is shown in 
Fig.~\ref{detfreqcomp} for various values of $\sigma_{\mathrm{RV}}$. For comparison 
we also include 
the completeness limits for rocky planets (Earth-like composition) and planets with 
extended envelopes (Neptune-like composition). 

In the best case scenario where $\sigma_{\mathrm{RV}}=1$ m s$^{-1}$, we are 
50\% complete for detecting 
potentially habitable planets with \nobs{} $\sim 50$. A maximum detection 
completeness of $\sim 85$\% is then realized at \nobs{} $\sim 200$ although it is 
difficult to place an exact boundary on where the maximum completeness is reached given 
the comparatively large Poisson errors on the potentially habitable planet 
completeness curve. If $\sigma_{\mathrm{RV}}$ is doubled, the maximum detection 
completeness falls to $\sim 70$\%. 
These complete curves for potentially habitable planets are very similar to the 
equivalent curves for the full sample of planets (c.f. Fig.~\ref{detfreq}). We 
therefore conclude that we are as efficient at detecting potentially habitable 
planets as we are to detecting all planets.

\begin{figure}
\centering
\includegraphics[scale=.6]{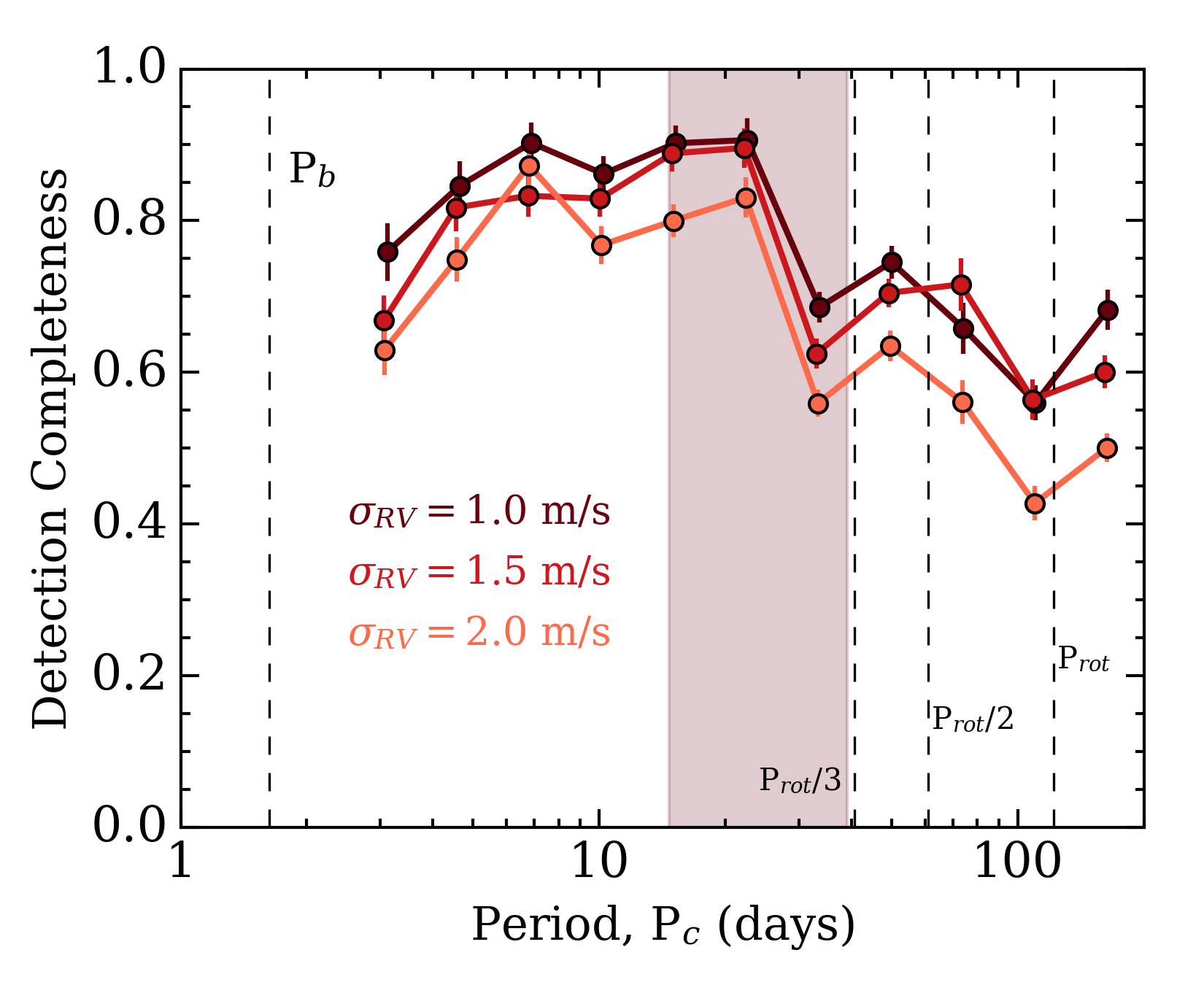}
\caption{Detection completeness as a function of orbital period and RV measurement
uncertainty. The vertical dashed 
lines highlight the orbital period of GJ 1132b ($P_b=1.62893$ days), the stellar 
rotation period ($P_{\mathrm{rot}}=122.31$ days), and its harmonics. The habitable zone is 
depicted in the shaded red region.}
\label{Pdetfreq}
\end{figure}

Comparing the potentially habitable completeness curves to the detection completeness 
of \emph{all} rocky planets in the sample, we observe that we are more sensitive to HZ 
rocky planets than to the total population of rocky planets. 
Fig.~\ref{Pdetfreq} reveals that the period range over which we are maximally 
efficient at detecting planets approximately corresponds to the inner edge of the habitable 
zone. At both shorter and longer orbital periods, the detection completeness falls 
significantly compared to the domain of maximum completeness approximately 
spanning $P \in [5, 20]$ days, which spans the inner HZ edge. Recall that we have intentionally 
avoided periodicities close to the stellar rotation period and its low-order harmonics 
including $P_{\mathrm{rot}}/2$; the effective periodicity of the RV jitter 
from the $FF'$ method. 
We draw attention to these periods because the detection completeness 
of planets with orbital periods close to these periodicities is artificially reduced. However, 
the fraction of planets within only $\pm 2$\% of either $P_{\mathrm{rot}}$ or $P_{\mathrm{rot}}/2$ 
is small such that the maximum 
detection efficiency still corresponds to the peak in Fig.~\ref{Pdetfreq}. 

Returning to Fig.~\ref{detfreqcomp} we also note that, unsurprisingly, we are far more 
sensitive to large ($r_{p,c} > 1.6$ R$_{\oplus}$), Neptune-like planets given their 
corresponding large masses compared to the Earth-like planets in the sample (c.f. 
Fig.~\ref{planetdist}). Regardless of $\sigma_{\mathrm{RV}}$, the detection of 
Neptune-like planets is always at least 20\% more complete than for Earth-like 
planets for all values of \nobs{.}

\subsection{Period-Mass Plane}
\begin{figure}
\centering
\includegraphics[scale=.45]{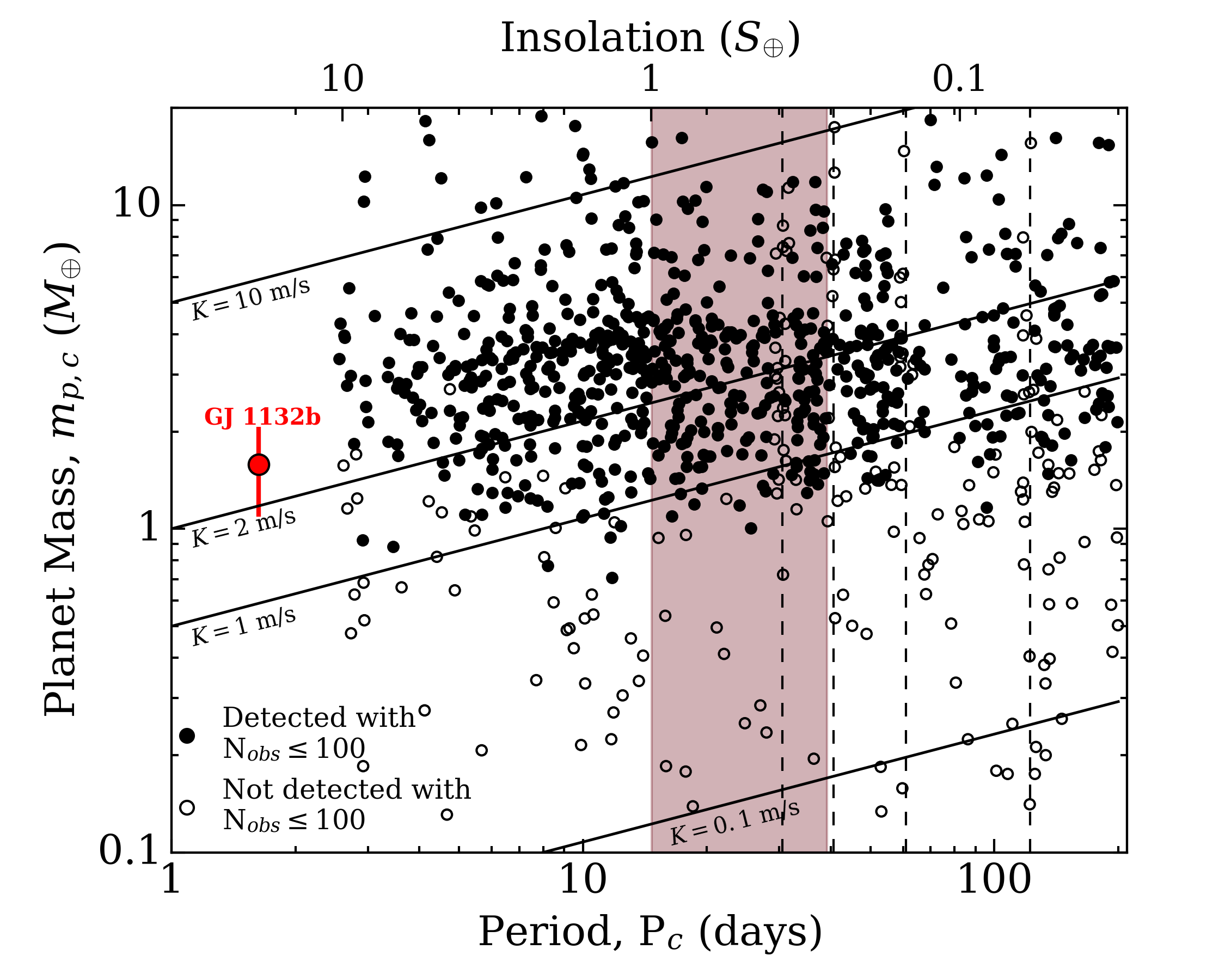}
\caption{The planetary mass-period/insolation plane of our full MC planet sample with 
$\sigma_{\mathrm{RV}}=1$ m s$^{-1}$. Planets 
which are detected with $\leq 100$ RV measurements are shown in black 
whereas planets which remain undetected are shown in white. Contours of constant 
RV semiamplitude are over-plotted as sold black lines. The vertical dashed lines indicate the 
rotation period of GJ 1132 and its low-order harmonics. The habitable zone is depicted in 
the shaded red region. GJ 1132b is highlight as the red circle.}
\label{MpS}
\end{figure}

To aid in the interpretation of the results from our MC simulations we plot the 
distribution of simulated planets in the period-mass plane in Fig.~\ref{MpS} 
and indicate which planets are detected with \nobs{} $\leq 100$. We do this for the 
case of zero systematic uncertainty; $\sigma_{\mathrm{RV}}=1$ m s$^{-1}$. For reference, the 
detection completeness for this full sample and the subset of potentially habitable 
planets is $75 \pm 3$\% and $71 \pm 9$\% respectively.

Notable features include the significant fraction of planets detected with $K_c \geq 1$ 
m s$^{-1}$ including a population of sub m s$^{-1}$ planets with orbital periods 
$\gtrsim 10$ days. There is also a distinct dearth of planets with $P_c \lesssim 2.5$ 
days. This is a consequence of the existence of GJ 1132b with $P_b = 1.62893$ days and 
the dynamical stability arguments imposed on each simulated planetary system. 

Also recall that we purposely ignore periodicities in the Lomb-Scargle periodogram 
analysis which lie close to the stellar rotation period and its low-order harmonics. 
Such planets lie close to the vertical dashed lines in Fig.~\ref{MpS} but appear to 
be embedded in regions where their nearest neighbours are detected. Therefore, if the 
stellar rotation period were to vary slightly from $P_{\mathrm{rot}} = 122$ days, 
then these 
planets would likely be detected potentially increasing the detection completeness. 
Some of these disregarded planets also appear to lie within the habitable zone, 
making our detection completeness of potentially habitable planets around GJ 1132 
more favorable.

\section{Discussion} \label{discussion}
\subsection{Applicability to Other Systems} \label{tess}
Our work regarding the effort required to recover additional planets around slowly 
rotating M-dwarfs such as GJ 1132  
is applicable to other systems containing at least one transiting planet such 
as those which will be discovered with \emph{TESS} \citep{ricker14, sullivan15}. 
The condition of slow rotation is required to focus on systems with similarly low 
noise statistics from stellar jitter (see Sect.~\ref{jitter}).

Numerous studies of the rotation period distribution of M-dwarfs  
($T_{\mathrm{eff}} \le 3800$ K) have concluded 
that two populations exist corresponding to fast and slow rotators 
\citep[e.g.][]{irwin11, mcquillan13a, mcquillan14, newton16a}. 
This dichotomy among M-dwarfs is 
posited to arise from magnetic braking; angular momentum loss by strongly magnetized 
stellar winds \citep{reiners12b}. This implies that the rotation period of these low-mass, 
convective stars might be used to infer stellar ages via gyrochronology 
\citep{barnes03, barnes07}. Indeed the empirical distribution of M-dwarf rotation periods 
is in good agreement with the inferred distribution derived from gyrochronology 
\citep{mcquillan14}. By this logic, 
GJ 1132 is an old ($\sim 5$ Gyrs) star which belongs to a large population of slowly 
rotating \citep[$P_{\mathrm{rot}} \gtrsim 40$ days;][]{newton16a} M-dwarfs 
with correspondingly low RV jitter levels of just a few \mps{} or comparable to
the demonstrated and anticipated stability limit of state-of-the-art 
velocimeters.

From the rotation period distribution of \cite{newton16a}, roughly 40\% of 
\emph{TESS} M-dwarfs with planet 
candidates are expected to exhibit slow rotation and correspondingly 
low RV jitter levels similar to 
GJ 1132. For these systems, which will require RV follow-up observations to 
characterize planet masses and search for additional non-transiting planets, 
their orbital periods will be well-characterized by the transit observations
thus allowing the detection completeness to additional small planets around
the star to not differ significantly from what we have shown in this study.
This is because the orbital period and time of mid-transit for the
transiting planet candidates 
will be well-characterized by \emph{TESS} implying that the keplarian signal of the 
candidate can be removed from a RV timeseries with relative ease once its mass and 
orbital elements have been adequately constrained (see Sect.~\ref{sect:massdetsig}). 
The search for additional planets 
then proceeds identically to the investigation presented in this paper.

Indeed the applicability of our detection completeness calculations
is only true for slowly rotating M-dwarfs which are sufficiently bright such that
a RV stability limit of $\sigma_{\mathrm{RV}} \lesssim 2$ \mps{} can be achieved.
For the upcoming near-IR spectropolarimeter \spirou{,}
$\sigma_{\mathrm{RV}} \sim 1$ \mps{} is expected to be readily achieved 
for stars brighter than $\sim 9.5$ in the $J$ band. \emph{TESS} is predicted
to find $\sim 50$ such stars with transiting planets \citep{sullivan15}.
Other current and upcoming near-IR
spectrographs are anticipated to be able to achieve similar stability limits.
Furthermore, from the full \emph{TESS} survey predictions 
of \cite{sullivan15}, and the empirical stellar rotation period distribution as a 
function of stellar mass, 
we expect our detection completeness curves (Figs.~\ref{detfreq}, \ref{detfreqKs}, and 
\ref{detfreqcomp}) to 
be relevant to $\sim 300$ M-dwarfs from the \emph{TESS} catalogue. 
Although, special considerations will be required to observe stars
with $9.5 \lesssim J \lesssim 11$ if $\sigma_{\mathrm{RV}} \lesssim 2$ \mps{} is to be
achieved. 
Optimistically, this number may even represent a lower limit given the apparent dearth 
of planets around rapidly rotating stars \citep{mcquillan13b} implying that the population 
of detected \emph{TESS} candidates may preferentially exist around slow rotators.

However, a notable concern exists for potentially habitable planets around M-dwarfs 
wherein 
the orbital periods corresponding to the HZ may be close to the stellar rotation period and 
its low-order harmonics thus making it difficult to detect such planets \citep{vanderburg16}. 
With that in mind, \cite{newton16b} argued that 
M-dwarfs with spectral classes M4-M6 
($0.1 \lesssim M_s/$M$_{\odot} \lesssim 0.25$) 
represent the best possible candidates for finding HZ planets around cool stars as their HZ 
limits\footnote{The habitable zone period limits for M4V: $\sim 21-55$ days and for M6V: 
$\sim 6-17$ days.} lie within the 
peaks of the M-dwarf rotation period distribution.

\subsection{Detecting Planets Around Active Stars Using Gaussian Processes} \label{sect:rapid}
As we have noted, this study is most applicable to cases of known transiting systems around 
M-dwarfs with rotation periods $\gtrsim 40$ days. These systems are enticing for the search 
for small planets because the rotationally-induced RV jitter is expected to be 
`manageable'. In Sect.~\ref{sect:rvjm} 
we showed that when the RV jitter amplitude is $>\sigma_{\mathrm{RV}}$, 
that a GP trained on the star's light curve can effectively model the jitter and bring the 
rms of the residuals (RVs minus jitter model) to the level of $\sim \sigma_{\mathrm{RV}}$ (c.f. 
Fig.~\ref{fig:gpjitter}). 

However there exists a substantial fraction of M-dwarfs 
which belong to either a kinematically young population or are among the latest M-dwarfs 
which can remain magnetically active even with $P_{\mathrm{rot}} > 40$ days \citep{west15}. 
In these cases, the search for 
planets may not proceed identically to the methodology of this study due to the large amplitude 
RV jitter. Following the work of a number of other authors 
\citep[e.g.][]{haywood14, rajpaul15} 
here we show how a Gaussian process, trained on an activity diagnostic 
timeseries, can still be used to model the jitter from active regions 
and detect planets with $K$ much less than the stellar RV jitter signal. Accurate jitter modelling
in rapidly rotating M-dwarf systems also facilitates the detection of the Rossiter-McLaughlin
effect from small planets thus providing a direct measure of the angular momentum evolution
in such systems \citep{cloutier16}.

In particular, we consider a case identical to the GJ 1132 planetary system 
($P_b\sim 1.63$ days, $K_b = 2.69$ \mps{)} 
but we decrease the stellar rotation period by more than an order of magnitude to 
$P_{\mathrm{rot}} = 2$ days and correspondingly increase the amplitude of the star's photometric 
variability to a more suitable value of $10^4$ parts per million \citep[ppm;][]{newton16a}. 
In this case the consequential RV jitter amplitude is 
$\sim 60$ \mps{} \citep{aigrain12} instead of $\sim 8$ \mps{.} 
We construct a photometric timeseries from 
a simple sinusoid with the aforementioned period and amplitude, 
and a white-noise term with a typical dispersion of 1000 ppm \citep{newton16a}. 
We then derive the expected RV jitter timeseries 
from the light curve using the analytical prescription of \cite{aigrain12}. A synthetic 
RV timeseries is then created via the sum of the stellar jitter, derived from the synthetic 
light curve, and the GJ 1132b keplarian signal. We assume a fixed RV uncertainty of 
$\sigma_{\mathrm{RV}} = 1$ \mps{} for this exercise.

To model the effect of active regions 
and attempt to remove it and search for any planetary signals 
in the residuals, the methodology identical to what was used in Sect.~\ref{sect:rvjm} 
is adopted. In summary, we train a quasi-periodic 
Gaussian process (GP) on the stellar light curve. The marginalized posterior probability
distributions functions 
for the GP hyperparameters are derived using MCMC which are then used to compute a photometric 
model from which the RV jitter model is derived. This method of training the GP jitter model 
on an activity diagnostic timeseries and subsequently using the hyperparameter priors to model 
the observed RV with a GP + keplarian model has been successfully used on systems 
with jitter levels of a few to $\sim 10$ times greater than the planetary signal 
\citep{baluev13, haywood14, grunblatt15}, as is true in our current test case.

\begin{figure*}
\centering
\includegraphics[scale=.65]{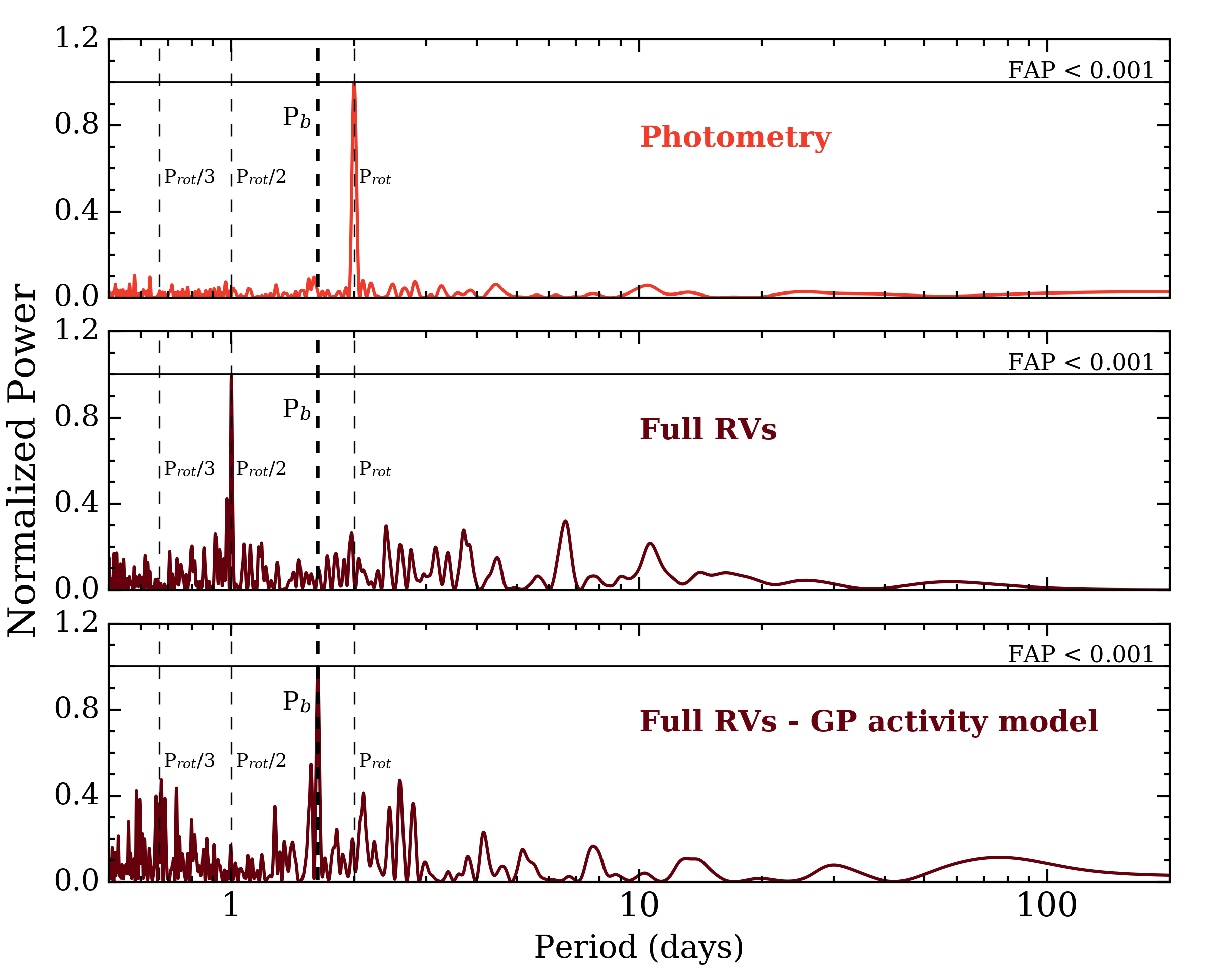}
\caption{\emph{Upper panel}: Lomb-Scargle 
periodogram of the synthetic photometric light curve 
($P_{\mathrm{rot}}=2$ days)
\emph{Middle panel}: Lomb-Scargle periodogram of the raw RV timeseries 
containing contributions from 
active regions and a planet with an orbital period distinct from $P_{\mathrm{rot}}$ 
($P\sim 1.63$ days). \emph{Lower panel}: the Lomb-Scargle 
periodogram of the RV residual timeseries after the subtraction of the mean 
Gaussian process activity model trained on the photometry. Upper limits on the 
false-alarm probability of 
significant peaks are highlighted by the horizontal solid 
lines. Vertical dashed lines highlight the injected planetary orbital 
period and the stellar rotation period plus its first and second harmonic periods.}
\label{periodogram}
\end{figure*}

The best evidence for detecting the planet following the removal of our MAP jitter model comes 
from a Lomb-Scargle periodogram given that we know the injected planet's orbital period. 
The upper panel of Fig.~\ref{periodogram} shows the periodogram of the photometric 
timeseries with an unambiguous peak at the 2 day stellar rotation period (FAP $< 0.1$\%). 
This should be obvious given the imposed sinusoidal nature of the synthetic light curve. 
The middle panel then shows the periodogram of the raw RV timeseries 
(active regions+planet+noise) with a significant peak at the first harmonic of 
the stellar rotation period; $P_{\mathrm{rot}}/2$. 
We expect to see the strongest periodicity at $P_{\mathrm{rot}}/2$ if the RV 
variation is strongly affected by the flux effect (see Sect.~\ref{sect:rvjm}) 
from active regions, as is the case for 
rapidly rotating stars, because this effect scales linearly with the first time 
derivative of the light curve. Because of the strong jitter signal compared to the 
planetary signal, the periodogram power at $P_b$ 
is well beneath the noise in this timeseries. 
Therefore we clearly do not detect the injected planetary signal when the stellar 
jitter is left untreated.

Once we remove the GP jitter model from the observed RVs, 
the Lomb-Scargle periodogram exhibits a new distinct peak at $P_b\sim 1.63$ days 
(FAP $< 0.1$\%; lower panel Fig.~\ref{periodogram}). 
This is by far the strongest peak in the residual timeseries 
and leads to the detection of the periodic planetary signal with high significance. This 
test further demonstrates the effectiveness of using Gaussian processes to detect low-amplitude 
planetary signals around active stars in RV if additional, activity-sensitive 
timeseries are obtained either contemporaneously or in quick succession to RV observations 
in order to avoid any 
discrepancies that may arise from long-timescale activity cycles.

\subsection{Improving the Mass Estimate of GJ 1132b with Large \nobs{}} \label{sect:massdetsig}
\begin{figure}
\centering
\includegraphics[scale=.5]{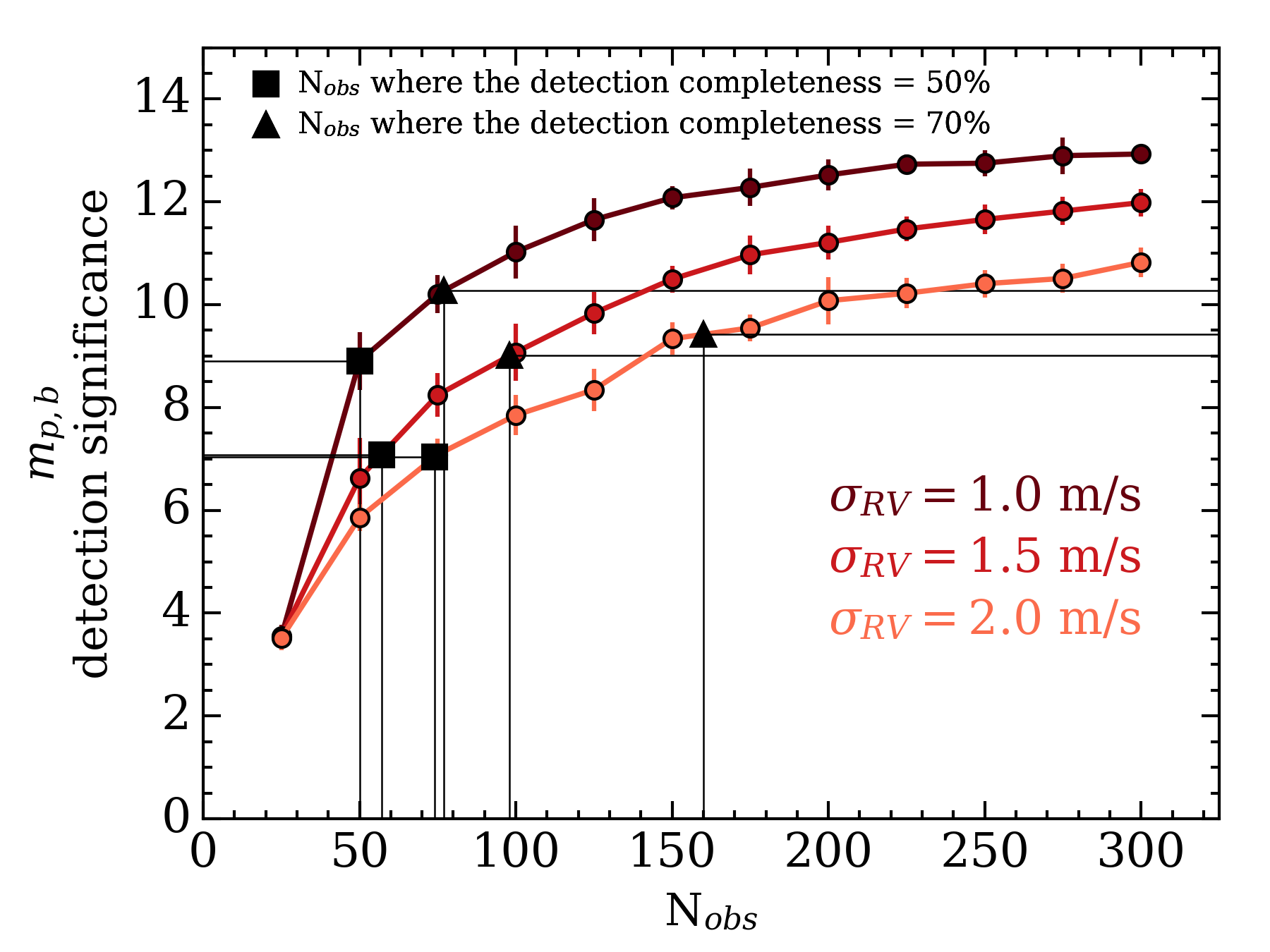}
\caption{The detection significance of the GJ 1132b planet mass of various RV 
measurement uncertainties. Errorbars result from the 
dispersion in values of $m_{p,b}$ when various window functions are used. The value of 
\nobs{} for which the detection completeness of all planets is 50\% (\emph{squares})
and 70\%  (\emph{triangles}; see Fig.~\ref{detfreq}) 
is depicted with \emph{vertical lines} for each value of $\sigma_{\mathrm{RV}}$.}
\label{harpsmcmc}
\end{figure}

A natural consequence of obtaining large values of \nobs{} to detect additional planets 
around GJ 1132 is that the semiamplitude $K_b$ can be better constrained in the process. 
For example, using just the 25 published RV measurements \citepalias{berta15}, 
the semiamplitude and planet mass $m_{p,b}$ are detected at a level of $\sim 3.4\sigma$. 
Here we compute the planet mass detection significance of GJ 1132b for increased values 
of \nobs{.} To avoid spurious effects on the planet mass measurement brought on by specific window 
functions, we use multiple draws of orbital phase to create multiple timeseries. 
We then fit the dataset 
with a single keplarian model using MCMC with the orbital period and time of mid-transit 
constrained by narrow uniform priors based on the uncertainties listed in Table~\ref{gj1132table}. 
The prior of $K_b$ is uniform on $[0,15]$ \mps{.} The $16^{\mathrm{th}}$ and 
$84^{\mathrm{th}}$ percentiles of the $K_b$ posterior distribution are used to compute its 
uncertainty 
which is then used along with the relevant values from Table~\ref{gj1132table} to compute 
the planet mass, its uncertainty, and hence its detection significance. 
The results are shown in Fig.~\ref{harpsmcmc}.

For \nobs{} $=25$ the detection significance is independent of $\sigma_{\mathrm{RV}}$ because 
the dataset only includes the published radial velocities whose noise properties are fixed. 
When $\sigma_{\mathrm{RV}}=1$ m 
s$^{-1}$, a bona fide mass detection of 5$\sigma$ ($10\sigma$) is achieved for GJ 1132b with \nobs{} 
$\sim 32$ ($\sim 70$).  
When $\sigma_{\mathrm{RV}}$ is increased to 2 \mps{,} 
a $5\sigma$ ($10\sigma$) mass detection requires \nobs{} $\sim 40$ ($\sim 200$). 

Fig.~\ref{harpsmcmc} also shows the mass detection significance obtained after a sufficient 
number of RV observations have been obtained in order to reach a 50\% and 70\% planet detection 
completeness. For instance, for $\sigma_{\mathrm{RV}}=1$ \mps{,} when we are 50\% (70\%) complete to 
additional planets in the system, the mass of GJ 1132b can be measured at the $\sim 9\sigma$ 
($\sim 10\sigma$) level. Similarly for $\sigma_{\mathrm{RV}}=2$ \mps{,} 
when we are 50\% (70\%) complete to 
additional planets in the system, the mass of GJ 1132b can be measured at the $\sim 7\sigma$ 
($\sim 9\sigma$). This assumes that the signal from additional planets have either a small RV 
signal or that they can be modelled 
with sufficient accuracy such that the residual rms is not large compared to 
the RV measurement uncertainty.

\subsection{Effect of Varying the Mass-Radius Relationship} \label{mrrel}
Throughout this work we have assumed a simplistic mass-radius relationship which 
assigns a bulk density equivalent to that of the Earth for small planets 
($r_p \leq 1.6$ R$_{\oplus}$) and the bulk density of Neptune for larger planets. This 
adopted relationship is not intended to be representative of the empirical distribution 
of planet masses and radii given that nature appears to present us with a range 
of planetary radii for given a mass \citep{seager07, zeng13}. Rather, we chose this 
relationship because of its simplicity and to remain agnostic regarding the exact form 
of the relationship between planetary mass and radius. Indeed many deterministic and 
probabilistic relations have been proposed. 

However, it is worth noting how our detection completeness changes if a different 
mass-radius relation is adopted. We test this using the two-regime fit from \cite{weiss14} 
(hereafter \citetalias{weiss14}) which 
allows the bulk density of rocky planets to increase linearly with $r_p$ for $r_p < 1.5$ 
R$_{\oplus}$. The relation then switches to a $m_p \propto r_p^{0.93}$ powerlaw for larger 
planets up to 4 R$_{\oplus}$. Each regime has an intrinsic scatter about the best-fit model 
which makes the relationship probabilistic. We use the model parameters reported in 
\citetalias{weiss14} to convert our sample of planet radii to masses and re-compute the 
detection completeness using this new sample of planet masses. 

\begin{figure}
\centering
\includegraphics[scale=.6]{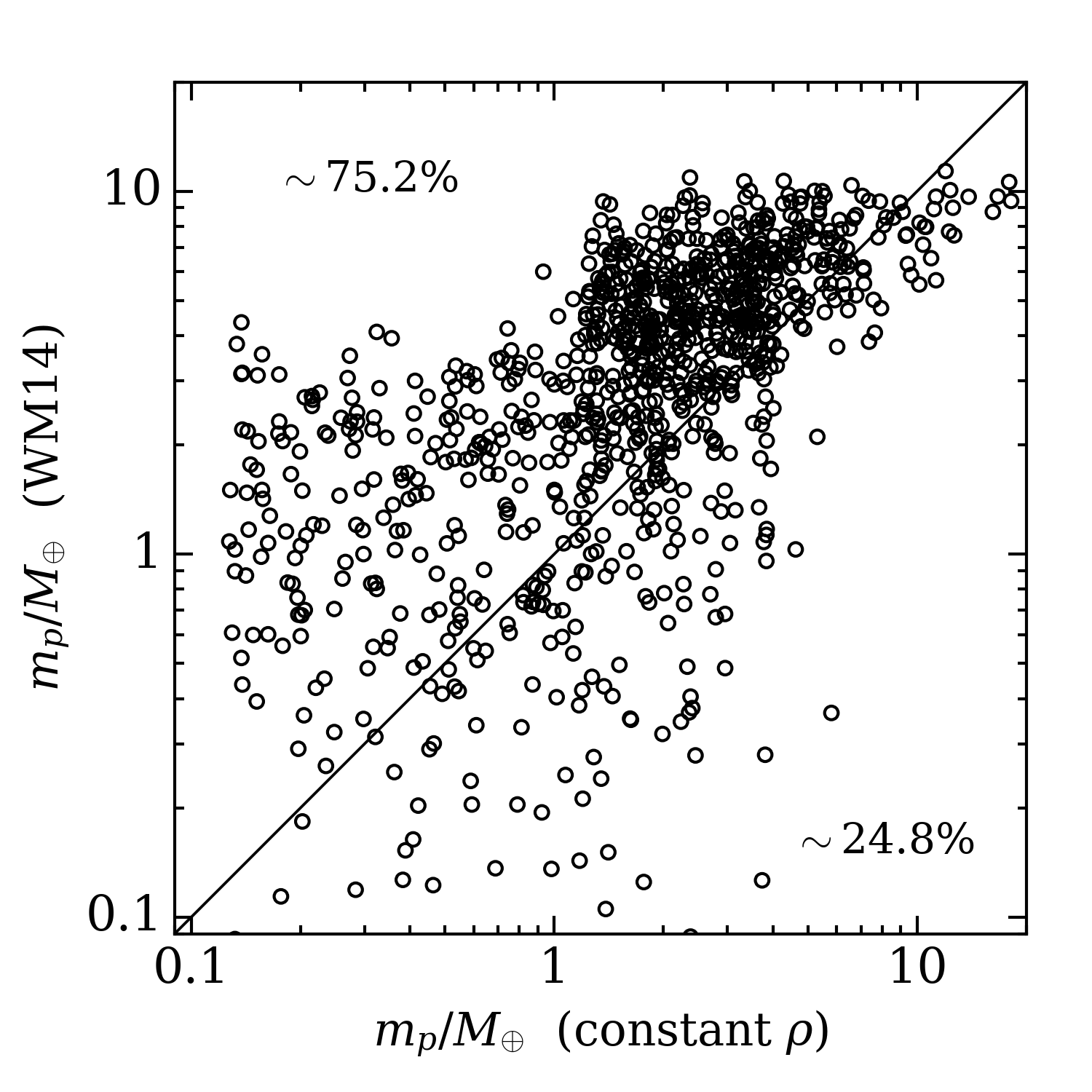}
\includegraphics[scale=.6]{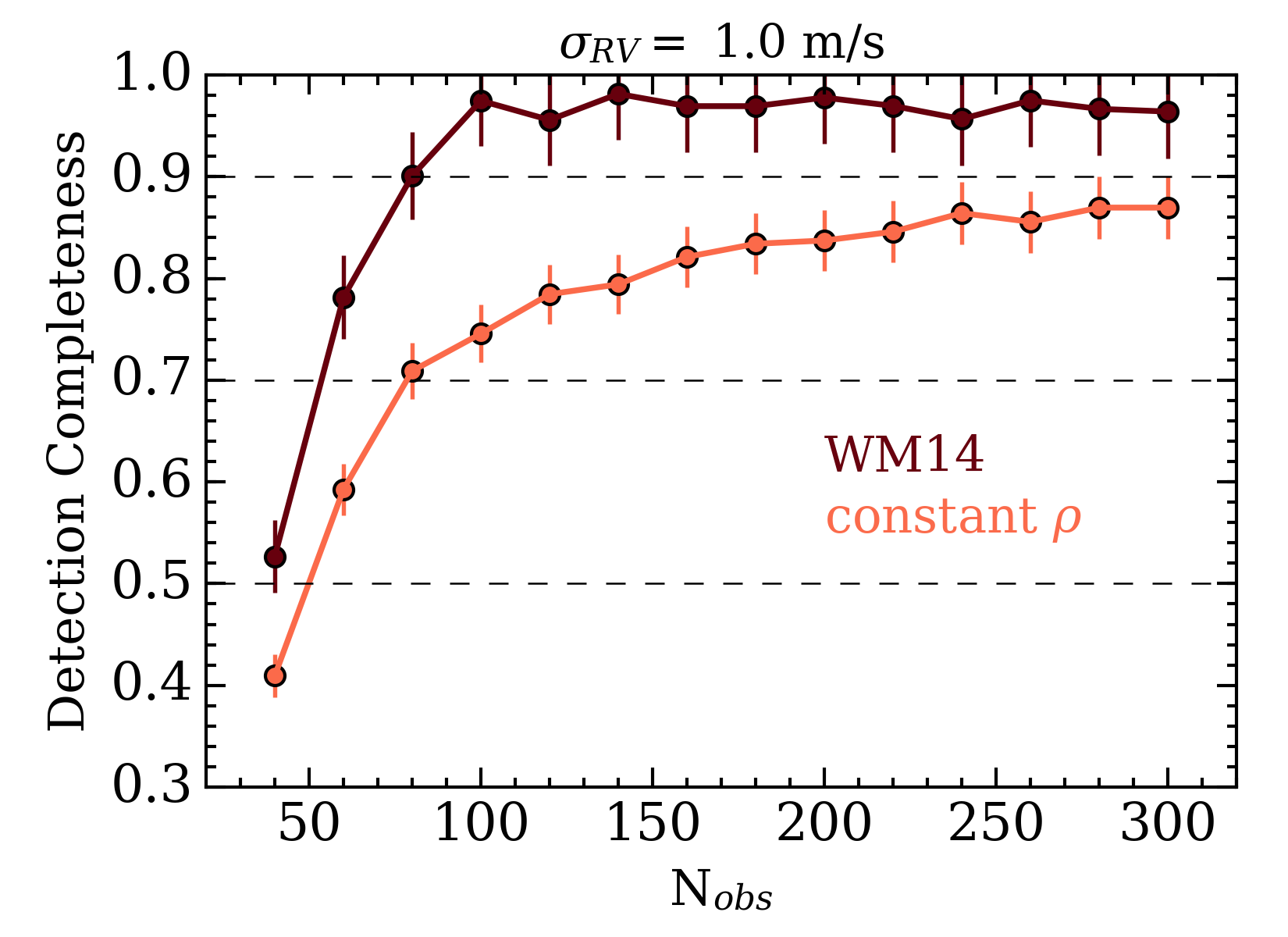}
\caption{\emph{Top panel}: the masses predicted by two planetary mass-radius relations: 
i) constant bulk density equal to either Earth or Neptune depending on the planet size; 
the relation used for the majority of this work or ii) the two-regime fit to 
the empirical planet distribution from \cite{weiss14}. \emph{Bottom panel}: the 
detection of planets around GJ 1132 for the two planetary mass-radius relationships.}
\label{mrrelplot}
\end{figure}

The discrepancies that arise from adopting different mass-radius relationships are illustrated 
in Fig.~\ref{mrrelplot}. Firstly, taking a subset of 1000 planet radii from our MC sample, 
we compute the planet masses using each relationship 
and plot the resulting masses against each other in the top panel of Fig.~\ref{mrrelplot}. It 
is clear that the \citetalias{weiss14} planet masses are frequently greater than the resulting mass 
when assuming a constant bulk density. For a given $r_p$, \citetalias{weiss14} returns 
a larger $m_p$ than our adopted mass-radius relationship in $\sim 75$\% of cases. This translates 
into a detection completeness which is greater than that which was previously obtained in our 
full MC simulation. The lower panel of Fig.~\ref{mrrelplot} demonstrates that the detection 
completeness increases by $\sim 10-25$\% if we adopt the \citetalias{weiss14} mass-radius 
relation 
thus, on average obtaining larger $m_p$ from our sample of $r_p$. Therefore, the fiducial 
mass-radius relation used in our study represents a conservative case as 
adopting a different mass-radius relation based on the empirical distribution of planets will 
yield higher mass planets thus making them easier to detect in RV.

\subsection{Detection Completeness for Planet Population Statistics}
We noted in Sect.~\ref{tess} that the planet detection completeness curves shown in 
Figs.~\ref{detfreq}, \ref{detfreqKs}, and \ref{detfreqcomp} will be applicable to a significant number 
($\sim 280$) of \emph{TESS} systems with planet candidates. In other words, the detection 
completeness of \emph{nearly edge-on}, 
small planets around slowly rotating M-dwarfs has, to a good approximation, 
been computed using GJ 1132 as a test case. The detection completeness as a function of 
RV measurements is useful when attempting to infer global 
properties of a planet 
population. In particular, for the aforementioned \emph{TESS} systems, 
our detection completeness, 
coupled with the results of RV follow-up observations contains information on the 
number of planets in a given system. This is true given that the known occurrence rates 
of planets around M-dwarfs was naturally folded into our MC simulations and therefore 
into our detection completeness curves.

\section{Summary and Conclusions} \label{conclusion}
This study aims at investigating the detection completeness of small planets around 
slowly rotating M-dwarfs when the planetary system is orientated nearly edge-on 
using precision radial velocity (RV) measurements. 
This will be the case for a large subset of \emph{TESS} candidate systems which 
will contain at least one transiting planet requiring follow-up observations to 
characterize the planet's mass and to search for additional planets in the system 
not seen to transit. 
In practise, we perform this calculation using GJ 1132 as a fiducial test case. Using 
existing photometric and RV data of this nearby 
M-dwarf and its short-period rocky companion GJ 1132b, we simulate the expected 
RV signal of the star at future epochs and compute the detection 
completeness of additional planetary companions in the system under realistic 
observing conditions. Our main results are as follows.

\begin{itemize}
\item Using a non-parametric Gaussian process to model the star's photometric variability, 
we measure a stellar rotation period  $P_{\mathrm{rot}} \sim 122$ days, consistent 
with previous estimates \citep{berta15}, and predict that the associated level of 
RV jitter to be $\sim 8$ \mps{.} A Gaussian process model is used to model the jitter 
down to a residual rms value comparable to the RV stability limit of 1 \mps{.}
\item We run a intensive suite of Monte-Carlo simulations which computes the expected 
RV signal from stellar jitter and small planets, sampled from their known occurrence 
rates, and quantifies the planet detection completeness as a function of the 
number of RV measurements.
\item We find that the detection completeness of 
all additional planets (excluding the known transiting planet), 
is $\sim 80-85$\% and is achieved with $\sim 200$ measurements for a nominal RV 
measurement uncertainty of $\sigma_{\mathrm{RV}} = 1$ \mps{.} 
Increasing $\sigma_{\mathrm{RV}}$ 
by a factor of 2 only worsens the detection completeness by $\sim 10-15$\%.
\item For a given number of measurements and RV measurement uncertainty, 
the detection completeness of potentially habitable worlds is found to be 
consistent with the detection completeness to the full planet sample. 
\item Limits on the number of RV measurements required to recover 50\% of i) all potential 
planets in the system and ii) all potentially habitable planets with state-of-the-art 
instrumentation is \nobs{} $\sim 50$. 
\item If contemporaneous ancillary timeseries, such as photometry, can be obtained, 
then the use of Gaussian processes are a powerful tool of modelling stellar jitter in active 
stars and detecting small amplitude planets.
\end{itemize}

Our detection completeness curves may also be applied to other slowly rotating 
M-dwarf systems containing 
at least one transiting planet such as those which will be found with \emph{TESS}. 
Knowledge of planet detection completeness as a function of the number of RV 
observations is incredibly useful for optimizing observing strategies aimed at 
the efficient detection of large samples of exoplanets. 
For cases in which planetary systems exist 
around active (often rapidly rotating) M-dwarfs, we recommend the use of 
Gaussian processes to distinguish between jitter and dynamical signals if 
ancillary timeseries such as photometry, polarimetry, multi-wavelength 
RV measurements, and/or spectral activity indicators (e.g. $\log{R_{\mathrm{HK}}'}$ 
or H$\alpha$) can also be obtained. 

\acknowledgements
RC would like to thank Jo Bovy, Daniel Fabrycky, and Daniel Tamayo for useful
discussions 
during the preparation of this manuscript and to the Canadian Institute for 
Theoretical Astrophysics for use of the Sunnyvale computing cluster throughout 
this work. RC is supported in part by the National Science and Engineering
Research Council and a Centre for Planetary Sciences Graduate 
Fellowship. KM is supported by the National Science and Engineering Research Council. 
X. Delfosse acknowledges the support of CNRS/PNP (Programme national de 
plan\'{e}tologie), 
CNRS/PNPS (Programme national de physique stellaire) and of Labex OSUG@2020. 
RC would also like to thank Zach Berta-Thompson and David Charbonneau for access to
the GJ 1132 MEarth photometry. 
This paper makes use of data from the MEarth Project, which is a collaboration 
between Harvard University and the Smithsonian Astrophysical Observatory. The 
MEarth Project acknowledges funding from the David and Lucile Packard Fellowship 
for Science and Engineering and the National Science Foundation under grants 
AST-0807690, AST-1109468, and AST-1004488 (Alan T. Waterman Award), and a grant 
from the John Templeton Foundation.

\appendix
\section{Modelling the photometric variability of GJ 1132 with a Gaussian process} \label{appendixGP}
Gaussian process (GP) regression is an attractive method for modelling the 
complex stochastic processes contributing to stellar variability and naturally 
lends itself to the class of Bayesian inferential statistics. Thus it 
is extremely well-suited to the computation of model hyperparameter values and
uncertainties and
has recently been applied in the literature to the recovery of stellar rotation periods 
\citep[e.g.][]{angus15, mancini15, vanderburg15} and to the modelling of RV jitter 
from spectroscopic activity diagnostics \citep[e.g.][]{rajpaul15} or  
photometry of active stars \citep[e.g.][]{haywood14, grunblatt15}. 

Motivated by previous observational studies \citep[e.g.][]{grunblatt15, mancini15, vanderburg15} 
and the notion that photometric variations due to rotating active regions evolve in a
quasi-periodic 
(QP) manner, we adopt a QP covariance function, or kernel, to describe the correlation 
between pairs of measurements. Explicitly, the time-correlation 
between data points is modelled as 

\begin{equation}
k_{i,j} = a^2 \exp{\left[ - \frac{|x_i-x_j|^2}{2\lambda^2} -\Gamma^2 
    \sin^2{\left(\frac{\pi|x_i-x_j|}{P_{\mathrm{rot}}} \right)} \right]},
\label{cov}
\end{equation}

\noindent where $x_i$ is the independent measurement epoch of the $i^{\mathrm{th}}$ 
measurement where the indices $i,j=1,\dots,N$ and $N$ is the number of data 
points. This covariance function is parameterized by four GP hyperparameters: 
the amplitude of the correlations $a$, the exponential decay timescale $\lambda$, 
the coherence `scale' of the correlations $\Gamma$, and the timescale of the periodic term 
$P_{\mathrm{rot}}$ which is representative of the stellar rotation period when using the
QP GP regression model to model the star's photometric variability. As 
mentioned, this kernel function is commonly used to model stellar light curves in 
a non-parametric way and returns the stellar rotation period, an important measurable 
property of the star, as a by-product of sampling the GP hyperparameters. 

To sample the probability distribution functions of the GP hyperparameters we ran a 
Markov chain Monte-Carlo (MCMC) simulation using the MEarth photometric light 
curve (Fig.~\ref{fig:mearthphoto}), 
re-sampled in 4 day bins to reduce the simulation's computational expense ($\mathcal{O}(N^3)$). 
The MCMC simulation was run using the affine-invariant MCMC ensemble sampler 
\texttt{emcee} \citep{foremanmackey13} coupled with the fast GP package 
\texttt{george} \citep{ambikasaran14, foremanmackey15} 
which is used to perform the necessary 
matrix inversions (see Eq.~\ref{like}). Throughout the MCMC the acceptance 
fraction of the sampler is monitored and constrained to $\sim 25-50$\%. 
Furthermore, we insist that the length of each Markov chain is at least a few ($\sim 10$) 
autocorrelation times such that we obtain multiple independent samples of each 
hyperparameter's marginalized posterior probability distribution function. 

The logarithmic likelihood function to be optimized by the MCMC is 

\begin{equation}
\ln{\mathcal{L}} = -\frac{1}{2} \left( \mathbf{y}^T K^{-1} \mathbf{y} 
+ \ln{\mathrm{det} K} + N \ln{2 \pi} \right) 
\label{like}
\end{equation}

\noindent where the $N \times N$ covariance matrix $K$ is

\begin{equation}
K_{i,j} = \sigma_i^2 \delta_{i,j} + k_{i,j}.
\end{equation}

\noindent In our case the vector $\mathbf{y}$ contains the $N$ differential magnitude 
measurements with variance $\sigma^2_i$ for the $i^{\mathrm{th}}$ measurement which 
are assumed to be uncorrelated. The $\delta_{i,j}$ denotes the Kronecker delta. 

\begin{figure}
\centering
\includegraphics[scale=.36]{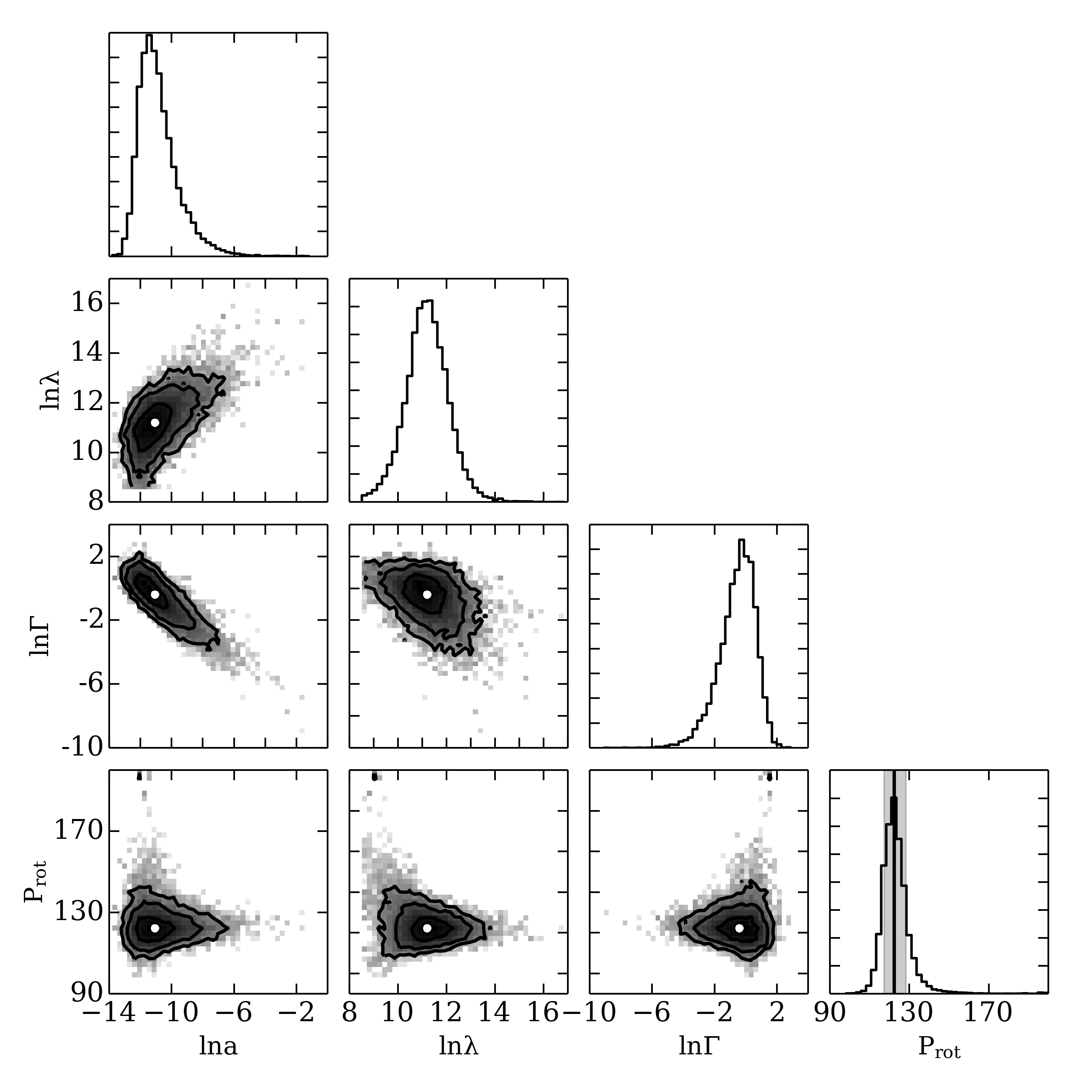}
\caption{Marginalized and joint posteriors for the quasi-periodic Gaussian process 
hyperparameters. Only $\lambda$ is constrained by a strict uniform prior of 
$\ln{\lambda} > 8.5$. All hyperparameters are well-constrained by the MEarth 
photometry including a measured stellar rotation of 
$P_{\mathrm{rot}} = 122.31^{+6.03}_{-5.04}$ days.} 
\label{post}
\end{figure}

The marginalized posterior probabilities for each hyperparameter are calculated 
up to a constant via the sum of Eq.~\ref{like} with a uniform logarithmic prior 
on each hyperparameter. All hyperparameters excluding $\lambda$ are effectively 
unconstrained by their broad uniform prior. From initial tests in which $\lambda$ 
is also left effectively unconstrained by a strict prior, a bi-modal $\lambda$ 
marginalized posterior distribution was observed which resulted in a corresponding 
value of $P_{\mathrm{rot}}$ not being well-defined. Each peak in the bi-modal 
$\lambda$ posterior was centered on a timescale greater than the expected 
stellar rotation period \citepalias[$P_{\mathrm{rot}} \sim 125$ days;][]{berta15} 
implying a long 
exponential decay timescale overlaid with the rotationally induced periodicity. By 
constructing two separate (uniform) priors on $\lambda$ that each spanned the 
values of one of the aforementioned peaks, we found that focusing on the first 
bi-modal peak (i.e. the shorter timescale peak) prevented a well-defined solution 
for $P_{\mathrm{rot}}$ from being found. Consequently, we limit 
$\ln{\lambda} > 8.5$ thus focusing on the second, longer timescale peak in the bi-modal
distribution. The result is 
well-defined values for all hyperparameters. A similar effect 
was found in \cite{rajpaul15} wherein restriction of the timescale $\lambda$ 
to values greater than some threshold prevented the convergence of 
$P_{\mathrm{rot}}$ to unphysical, short timescale `contortions'. The resulting 
marginalized and joint hyperparameter posteriors are shown in Fig.~\ref{post}. 

As an aside, we note that the MAP solution for 
$\lambda = 7.26 \times 10^4$ days is two orders of magnitude greater than the 
time baseline for the MEarth observations ($\sim 544$ days). Therefore, this 
timescale is not well-constrained by the data and additional photometry for the 
star is required to achieve a robust measurement of its value. Given the nature 
of the hyperparameter $\lambda$, the only remaining hyperparameter with a physical 
interpretation is $P_{\mathrm{rot}}$ whose marginalized posterior probability is 
shown in the bottom right panel of Fig.~\ref{post}. From its posterior 
we measure the MAP value of the stellar rotation period to be 
$P_{\mathrm{rot}} = 122.31^{+6.03}_{-5.04}$ days where the quoted uncertainties are based on the 
$16^{\mathrm{th}}$ and $84^{\mathrm{th}}$ percentiles. This result from the GP regression 
analysis is consistent with the measured rotation of $\sim 125$ days from 
\citetalias{berta15} determined using the sine-wave fitting methodology of 
\cite{irwin11} and \cite{newton16a}. 

\section{Planet detection via Bayesian model selection} \label{appendixZ}
Here we discuss the Bayesian model selection framework and its application to detecting 
planets in our Monte-Carlo (MC) simulations.

Firstly, we denote a complete dataset by the vector $\mathbf{D}$ containing 
$N_{\mathrm{obs}}$ RV measurements each with an associated error and taken 
at a unique BJD. These data are assumed to be derived from 
one of three proposed models (or hypotheses) denoted $M_k$ where $k=0,1,2$ denotes 
the number of planets in the model. The model $M_0$ is referred to as the null 
hypothesis to which we can compare other models of increasing complexity, to test 
their validity. Each model 
has a corresponding set of unique model parameters $\mathbf{\Theta}_k$ with 
dimensionality $D$. For models with $k>0$, the set of model parameters 
$\mathbf{\Theta}_k$ includes each planets' orbital period, time of mid-transit, 
and semiamplitude. All orbits are assumed to be circular to simplify the modelling 
process. 

Bayes theorem tells us that 

\begin{equation}
\prob(\mathbf{\Theta}_k | \mathbf{D}, M_k) = \frac{
  \prob(\mathbf{D} | \mathbf{\Theta}_k, M_k) \cdot \prob(\mathbf{\Theta}_k | M_k)}
     {\prob(\mathbf{D} | M_k)},
\end{equation}

\noindent where $\prob(\mathbf{\Theta}_k | \mathbf{D}, M_k)$ is the posterior 
probability of the $M_k$ model parameters, 
$\prob(\mathbf{D}|\mathbf{\Theta}_k, M_k) \equiv \mathcal{L}$ is the likelihood of 
obtaining the observed data given a set of model parameters, $\prob(\mathbf{\Theta}_k|M_k)$ is 
the prior on the model parameters, and $\prob(\mathbf{D}|M_k) \equiv Z$ is a 
normalization 
factor equal to the evidence for the model $M_k$. In general, computation of model 
evidences is not required for parameter estimation only but is necessary for model 
comparison and selection in our MC simulations. 

To determine which model best describes a dataset one must calculate the marginal 
model posterior probabilities $\prob(M_k|\mathbf{D})$ for each model $M_k$. The 
ratio of these values for two competing models in known as the Bayes factor and 
is commonly used to infer which model is most favorable for describing a dataset. 
Therefore, for arbitrary models designated $M_m$ and $M_n$, the Bayes factor is 

\begin{equation}
R_{mn} \equiv \frac{\prob(M_m|\mathbf{D})}{\prob(M_n|\mathbf{D})} \label{Bfact}
\end{equation}

\noindent where,

\begin{equation}
\prob(M_k | \mathbf{D}) = \frac{\prob(\mathbf{D} | M_k) \cdot \prob(M_k)}{
  \sum_{k} \prob(\mathbf{D} | M_k) \cdot \prob(M_k)} \label{probM}
\end{equation}

\noindent is the probability that the model $M_k$ describes the data. The factor in the 
denominator of Eq.~\ref{probM} 
contains a summation over all model evidences considered and therefore 
cancels when computing Bayes factors using Eq.~\ref{Bfact}. This cancellation 
allows us to write 
the Bayes factor describing the evidence for a model $M_m$ compared to $M_n$ as 

\begin{equation}
R_{mn} = \frac{\prob(\mathbf{D}|M_m)}{\prob(\mathbf{D}|M_n)}
\frac{\prob(M_m)}{\prob(M_n)}.
\end{equation}

Throughout this study we fix the ratio $\prob(M_m)/\prob(M_n)$ to unity for all 
$m,n$ to ensure that we are not biased towards any particular number of 
keplarians. This is intended to representative of the GJ 1132 case wherein although we 
expect there to be additional planets in the system based on the high frequency of 
small planets around M-dwarfs \citep{dressing13, dressing15a}, 
increasing this ratio to $\prob(M_{k>1})/\prob(M_{k=1}) > 1$ would favor additional 
planets which may or may not actually be present. 
The model evidences are then obtained by integrating over the model's full parameter 
space: 

\begin{equation}
Z_k \equiv \prob(\mathbf{D} | M_k) = \int \mathcal{L} \cdot
\prob(\mathbf{\Theta}_k | M_k) \cdot \mathrm{d}^D \mathbf{\Theta}_k.
\label{inte}
\end{equation}

Evaluation of the multi-dimensional integral in Eq.~\ref{inte} is an intensive 
computational task. To evaluate this integral for each model $M_k$ 
with $k=0,1,2$, we use the \texttt{Multinest} 
\citep{feroz08, feroz09, feroz14} nested sampling algorithm 
given the logarithmic likelihood function 

\begin{equation}
\log{} \mathcal{L} = -\frac{1}{2} \sum_{i=1}^{N_{\mathrm{obs}}} \left( 
\frac{(y_i-\mathrm{model}_i)}{\sigma_i} \right)^2 - 
\log{\left(\frac{1}{\sigma_i^2} \right)} \label{loglike}
\end{equation}

\noindent and appropriately chosen model parameter priors. The values $y_i$ and $\sigma_i$ denote 
the observed RV and its uncertainty respectively while model$_i$ 
is the value of the model $M_k$ for the measurement $i$ determined by the 
parameters $\mathbf{\Theta}_k$.  

As discussed in Sect.~\ref{pdetect}, the $M_1$ model parameter priors are based on the 
measured values for GJ 1132b in Table~\ref{gj1132table} whilst the prior on the orbital 
period for any additional planet is based on any low false-alarm probability 
peak in a Lomb-Scargle periodogram $P_\mathrm{peak}$ of the RV residuals (GJ 
1132 b keplarian removed). Priors on $T0_c$ and $K_c$ are left broad and uniform; 
$T0_c \in [T0_b-P_{\mathrm{peak}}/2, T0_b+P_{\mathrm{peak}}/2]$ BJD and $K_c \in (0, 15]$ m 
s$^{-1}$. Detection of a second planet therefore requires a periodogram periodicity 
detection coupled with a favorable Bayes factor of the two planet model 
compared to the single planet model; $R_{21} > 20$ \citep{kass95}.

\bibliographystyle{apj}
\bibliography{refs}

\end{document}